\documentstyle[12pt]{article}
\normalbaselines

\begin{document}

\renewcommand{\thefootnote}{\fnsymbol{footnote}}

\ \\

\vspace{-2.cm}

\begin{flushright}

{\raggedleft SWAT/96/105\\
hep-lat/9604024\\[2.cm]}

\end{flushright}

\begin{center}
{\LARGE\baselineskip0.9cm 
The exact equivalence of the two-flavour 

strong coupling lattice Schwinger model

with Wilson fermions to a vertex model\\[2cm]}

{\large K. Scharnhorst\footnote[2]{E-mail:
k.scharnhorst @ swansea.ac.uk}
}\\[0.3cm]
{\small University of Wales, Swansea

Department of Physics

Singleton Park

Swansea, SA2 8PP, U.K.}\\[1.5cm]
\end{center}
\renewcommand{\thefootnote}{\arabic{footnote}}

\thispagestyle{empty}

\begin {abstract}
In this paper a method previously employed by Salmhofer to establish 
an exact equivalence of the one-flavour strong coupling lattice
Schwinger model with Wilson fermions to some 8-vertex model 
is applied to the case with two flavours. As this method is 
fairly general and can be applied to strong coupling QED
and purely fermionic models with any (sufficiently small) number of 
Wilson fermions in any dimension the purpose of the present study is 
mainly a methodical one in order to gain some further experience with it. 
In the paper the vertex model equivalent to the two-flavour strong
coupling lattice Schwinger model with Wilson fermions 
is found. It turns out to be some modified 3-state
20-vertex model on the square lattice, which can also 
be understood as a regular 6-state vertex model. 
In analogy with the one-flavour case, 
this model can be viewed as some loop model.\\
\end{abstract}

\newpage
The motivation for the present work derives from a paper by
Salmhofer \cite{salm} who demonstrated the exact equivalence 
of the one-flavour strong coupling lattice Schwinger model with Wilson
fermions to a 8-vertex model which can also be understood as a 
certain self-avoiding loop model. While strongly coupled gauge
theories with staggered fermions lead to pure monomer-dimer systems
\cite{ros} the consideration of Wilson fermions results in more 
complicated models, as the investigation of Salmhofer \cite{salm} demonstrates
by example. The particular equivalence established by him
has led to hitherto unfeasible computer studies  
\cite{gaus3}--\cite{gaus4} as well as to 
the application of certain analytical methods \cite{scharn} yielding
previously unknown information about the critical behaviour of
the one-flavour strong coupling lattice Schwinger model.
However, the method employed by Salmhofer \cite{salm} is not specific to
this model. It can equally be applied, under certain technical complications, 
to strong coupling lattice QED 
(in fact, to any gauge theory, with some modifications) 
and purely fermionic models with any 
(sufficiently small) number of Wilson fermions in any dimension.
It is therefore particularly interesting for models with an odd
number of fermion species which give rise to serious difficulties for
any numerical simulation so far. Furthermore, the vertex model
equivalence (which has to be found case by case) of certain lattice 
models of quantum field theory with Wilson fermions
possibly will allow to make use of cluster algorithms 
for their numerical study in order to reduce 
the computational effort required. Beyond this, the 
equivalence may also lead to new ideas for analytical investigations.\\

In this paper we extend the analysis of Salmhofer \cite{salm}
to the two-flavour strong coupling lattice Schwinger model. Beyond 
the specific result of the paper -- the vertex model equivalence for the 
model under consideration -- the main purpose of it is to collect some 
practical experience in dealing with models which have a 
larger number of Grassmann variables
per lattice site than the one-flavour Schwinger model with Wilson
fermions, which has four. The main difficulty is an algebraic one. While 
the study of Salmhofer \cite{salm} requires paper and pencil only,
to go beyond becomes possible only by using an appropriate formula 
manipulation program. In this study we are using a program for 
Mathematica \cite{math} which can efficiently handle Grassmann 
variables and which has been designed just for the present purpose.\\

Before going into technical details, let us shortly describe the 
method to be used in words \cite{salm}. The square lattice $\Lambda$ the 
two-flavour Schwinger model is defined on is divided into an even 
and an odd sublattice. Use will be made of the fact that 
effectively only fermionic degrees of freedom are left in the theory under 
consideration (In the strong coupling Schwinger model the bosonic 
gauge field degrees of freedom are integrated out exactly first, 
leaving the fermionic sector for further consideration.). 
First, on the (say) even sublattice $\Lambda_e$ the Grassmann 
integration is carried out. This is a local procedure (the reason for 
it being, that at most nearest-neighbour couplings are present in the action) 
which amounts to a completely algebraic problem due to 
the Grassmannian character of the integration. The remaining Grassmann 
variables live on the odd sublattice $\Lambda_o$ and certain graphical 
rules, i.e.\ vertices, can be assigned to their combinations in a fairly 
natural way. In the final step an analysis can then be made what structures 
(lattice clusters) built of these vertices the remaining Grassmannian 
integration on the odd sublattice leads to. Since for Wilson fermions the 
lattice is homogeneous the result can be described in terms of a 
vertex model.\\

We are now prepared to pursue the program for the two-flavour Schwinger
model with Wilson fermions (with Wilson parameter $r = 1$). For easier
reference, our notation follows closely that applied in \cite{salm}.
The partition function $Z_\Lambda$ of the model on 
the square lattice $\Lambda$ is given by 

\parindent0.em

\begin{eqnarray}
\label{B1}
Z_\Lambda &=& \int DU \prod_{i=1}^2 D\psi_i D\bar\psi_i\ \ {\rm e}^{-S}\ \ \ ,
\end{eqnarray}

where $D\psi_i D\bar\psi_i = \prod_{x\in\Lambda}\prod_{\alpha=1}^2
d\psi_{i,\alpha}(x)\ d\bar\psi_{i,\alpha}(x)$ 
denotes the multiple Grassmann integration on the lattice
($i$ is the flavour index.). The action $S$ is defined by

\begin{eqnarray}
\label{B2}
S &=& S_F + \beta S_G\ \ \ \ \ ,\\[0.3cm]
\label{B3}
S_F &=& \sum_{x\in \Lambda} \sum_{i=1}^2 \left( {1\over 2}
\sum_\mu \left(\bar\psi_i(x+ e_\mu)(1+\gamma_\mu)
U_\mu (x) \psi_i(x)\right.\right.\nonumber\\[0.3cm]
&&\ \ \ + 
\left.
\bar\psi_i(x)(1-\gamma_\mu) U_\mu^\dagger(x) \psi_i(x+ e_\mu)
\right) - M_i \bar\psi_i(x)\psi_i(x)\Bigg)\ \ \ \ \ ,
\end{eqnarray}

and $U_\mu =\exp{[-i A_\mu]}$, $\beta = 1/g^2$.
$S_G$ is the standard Wilson action and the hopping parameters
$\kappa_i$ are given by $\kappa_i = 1/2M_i$ (For the sake of 
generality we allow for different masses for each flavour here.).
In the strong (infinite) coupling limit $\beta = 0$ the gauge field 
integration can be performed exactly and one finds (cf.\ eq.\ (5)
in \cite{salm})

\begin{eqnarray}
\label{B4}
Z_\Lambda &=& \sum_{k_l \in \{0,1,2,3,4\}}
\ \int \prod_{i=1}^2 D\psi_i D\bar\psi_i\ \ 
\exp\left(\sum_{x\in \Lambda} \sum_{j=1}^2 M_j \bar\psi_j(x)\psi_j(x)\right)
\nonumber\\[0.3cm]
&&\hspace{-1.7cm}\times\ \prod_{l = (x,x+e_\mu)} {1\over k_l !^2}
\left[ 
\left(\sum_{m=1}^2 \bar\psi_m(x) T_\mu^{(-)}\psi_m(x+ e_\mu)\right)
\left(\sum_{n=1}^2 \bar\psi_n(x+ e_\mu) T_\mu^{(+)}\psi_n(x)\right)
\right]^{k_l} .
\end{eqnarray}

In the two-flavour Schwinger model, the occupation numbers $k_l$
cannot exceed 4 because on each lattice site only 8 Grassmann variables
are present. In fact, explicit calculation shows that by virtue
of the nilpotency of the Grassmann variables $k_l$ cannot exceed 2.
As in \cite{salm}, we choose $\gamma_1 =\sigma_3$, $\gamma_2 = \sigma_1$ 
($\sigma_i$ are Pauli matrices.)
and the projection operators $T_\mu^{(\epsilon)} = (1+\epsilon\gamma_\mu)/2$,
$\epsilon\in \{ -1,1\}$ then read $T_1^{(+)}={\rm diag}\{1,0\}$,
$T_1^{(-)}={\rm diag}\{0,1\}$, 
$T_2^{(\epsilon)}={1\over 2}{1\,\epsilon\choose\epsilon\,1}$.
It is now advantageous in order to simplify further results 
to define for each flavour combinations of the original $\psi$ fields
in which terms $T_2^{(\epsilon)}$ is diagonal (The index in front of 
the comma always is the flavour index, the one behind it the 
component index.)

\begin{eqnarray}
\label{B5}
\chi_i(x)&=&U \psi_i(x)\ =\ {1\over\sqrt{2}}
\left(\psi_{i,1}(x) + \psi_{i,2}(x),
\psi_{i,1}(x) - \psi_{i,2}(x)\right)^T\ \ \ ,
\\[0.3cm]
\bar\chi_i(x)&=&\bar\psi_i(x) U\ =\ {1\over\sqrt{2}}
\left(\bar\psi_{i,1}(x) + \bar\psi_{i,2}(x),
\bar\psi_{i,1}(x) - \bar\psi_{i,2}(x)\right)
\end{eqnarray}

with

\begin{eqnarray}
\label{B6}
U&=&{1\over\sqrt{2}}\left(
\begin{array}{cr}
1 & 1 \\ 1 & -1
\end{array}\right)
\ =\ U^{-1}\ \ \ .
\end{eqnarray}

\parindent1.5em

In the expression for the partition function (\ref{B4}) the 
integration on the even sublattice $\Lambda_e$ is now performed
first. As each lattice site $x$ supports 8 Grassmann variables 
terms which respect 

\parindent0.em

\begin{eqnarray}
\label{B7}
\sum_{l\ni x}\ k_l\ +\ s_x&=&4
\end{eqnarray}

can give a non-vanishing contribution only ($s_x$ is the power of 
$\sum_{i=1}^2 M_i \bar\psi_i(x)\psi_i(x)$ in eq.\ (\ref{B4}).). 
Explicit calculation now reveals that only terms with $s_x=0$ or
$s_x=4$ are non-vanishing.
A given set of $k_l$ on the lattice $\Lambda$ uniquely
determines a certain configuration contributing 
to the partition function. Below we give the results for all
local integrals related to a given lattice site $x\in\Lambda_e$ 
which allow non-vanishing contributions. 
The results have been arranged in a way most suitable for the 
further discussion. While 
in \cite{salm} the calculation for the one-flavour model
has been performed by means of paper and pencil, the two-flavour
case requires use of the computer and we have relied on a 
purpose-written Mathematica program \cite{math} to do the necessary algebra.
The graphical rules below have to be interpreted as follows: the black
dot denotes any point $x$ on the even sublattice $\Lambda_e$ 
at which the Grassmann integration
is performed, a dashed line means $k_l = 0$ while thin and thick 
lines stand for $k_l = 1$ and $k_l = 2$ respectively.
The first coordinate component of $x$ is understood to be the 
horizontal one while the second component runs vertically.\\

\parindent1.5em

There is one possible vertex with $s_x=4$:\\

\parindent0.em
Vertex 1
\nopagebreak

\vspace{5mm}
\nopagebreak

\unitlength1.mm
\begin{picture}(150,27)
\put(0,12){
\unitlength1.mm
\begin{picture}(15,15)
\linethickness{0.15mm}
\put(12,7.5){\line(1,0){1}}
\put(10,7.5){\line(1,0){1}}
\put( 8,7.5){\line(1,0){1}}
\put( 6,7.5){\line(1,0){1}}
\put( 4,7.5){\line(1,0){1}}
\put( 2,7.5){\line(1,0){1}}
\put(7.5,12){\line(0,1){1}}
\put(7.5,10){\line(0,1){1}}
\put(7.5, 8){\line(0,1){1}}
\put(7.5, 6){\line(0,1){1}}
\put(7.5, 4){\line(0,1){1}}
\put(7.5, 2){\line(0,1){1}}
\put(7.5,7.5){\circle*{1.5}}
\end{picture}  }
\put(0,30.5){
\parbox[t]{15cm}{
\begin{eqnarray}
\label{V1}
&\ \ =&\ \ \ \int\prod_{i=1}^2\prod_{\alpha=1}^2
d\psi_{i,\alpha}(x)\ d\bar\psi_{i,\alpha}(x)\ {1\over 4!}\ 
\left(\sum_{i=1}^2 M_i \bar\psi_i(x)\psi_i(x)\right)^4\nonumber\\[3mm]
&\ \ =&\ \ \ M_1^2 M_2^2 
\end{eqnarray} }}
\end{picture}

All further vertices have $s_x=0$ of course.\\

Vertex 2
\nopagebreak

\vspace{5mm}
\nopagebreak

\unitlength1.mm
\begin{picture}(150,20)
\put(0,5){
\unitlength1.mm
\begin{picture}(15,15)
\linethickness{0.15mm}
\put(12,7.5){\line(1,0){1}}
\put(10,7.5){\line(1,0){1}}
\put( 8,7.5){\line(1,0){1}}
\put( 6,7.5){\line(1,0){1}}
\put( 4,7.5){\line(1,0){1}}
\put( 2,7.5){\line(1,0){1}}
\linethickness{0.6mm}
\put(7.5,2){\line(0,1){11}}
\put(7.5,7.5){\circle*{1.5}}
\end{picture} }
\put(0,21){
\parbox[t]{15cm}{
\begin{eqnarray}
\label{V3}
&\ \ =&\ \ \ \ \
\bar\chi_{1,1}(x+e_2)\ \bar\chi_{2,2}(x-e_2)\ \bar\chi_{1,2}(x-e_2)
 \ \bar\chi_{2,1}(x+e_2)\nonumber\\ 
&&\ \times\ \chi_{1,2}(x+e_2)\ \chi_{2,1}(x-e_2)\ \chi_{1,1}(x-e_2)
\ \chi_{2,2}(x+e_2)\ \ 
\end{eqnarray}  }}
\end{picture}

Vertex 3
\nopagebreak

\vspace{5mm}
\nopagebreak

\unitlength1.mm
\begin{picture}(150,20)
\put(0,5){
\unitlength1.mm
\begin{picture}(15,15)
\linethickness{0.15mm}
\put(7.5,12){\line(0,1){1}}
\put(7.5,10){\line(0,1){1}}
\put(7.5, 8){\line(0,1){1}}
\put(7.5, 6){\line(0,1){1}}
\put(7.5, 4){\line(0,1){1}}
\put(7.5, 2){\line(0,1){1}}
\linethickness{0.6mm}
\put(2,7.5){\line(1,0){11}}
\put(7.5,7.5){\circle*{1.5}}
\end{picture} }
\put(0,21){
\parbox[t]{15cm}{
\begin{eqnarray}
\label{V2}
&\ \ =&\ \ \ \ \ 
\bar\psi_{1,1}(x+e_1)\ \bar\psi_{2,2}(x-e_1)\ \bar\psi_{1,2}(x-e_1)
 \ \bar\psi_{2,1}(x+e_1) \nonumber\\
&&\ \times\ \psi_{1,2}(x+e_1)\ \psi_{2,1}(x-e_1)\ \psi_{1,1}(x-e_1)
\ \psi_{2,2}(x+e_1)\ \ 
\end{eqnarray}  }}
\end{picture}

Vertex 4
\nopagebreak

\vspace{5mm}
\nopagebreak

\unitlength1.mm
\begin{picture}(150,20)
\put(0,5){
\unitlength1.mm
\begin{picture}(15,15)
\linethickness{0.15mm}
\put( 6,7.5){\line(1,0){1}}
\put( 4,7.5){\line(1,0){1}}
\put( 2,7.5){\line(1,0){1}}
\put(7.5,12){\line(0,1){1}}
\put(7.5,10){\line(0,1){1}}
\put(7.5, 8){\line(0,1){1}}
\linethickness{0.6mm}
\put(7.5,7.5){\line(1,0){5.5}}
\put(7.5,2){\line(0,1){5.5}}
\put(7.5,7.5){\circle*{1.5}}
\end{picture} }
\put(0,21){
\parbox[t]{15cm}{
\begin{eqnarray}
\label{V6}
&\ \ =&\ \ \ \ \ 
\bar\chi_{1,2}(x-e_2)\ \bar\psi_{2,1}(x+e_1)\ \bar\psi_{1,1}(x+e_1)\ 
\bar\chi_{2,2}(x-e_2)\nonumber\\
&&\ \times\ \chi_{1,1}(x-e_2)\ \psi_{2,2}(x+e_1)\ \psi_{1,2}(x+e_1)\ 
\chi_{2,1}(x-e_2)/4
\end{eqnarray}  }}
\end{picture}

Vertex 5
\nopagebreak

\vspace{5mm}
\nopagebreak

\unitlength1.mm
\begin{picture}(150,20)
\put(0,5){
\unitlength1.mm
\begin{picture}(15,15)
\linethickness{0.15mm}
\put(12,7.5){\line(1,0){1}}
\put(10,7.5){\line(1,0){1}}
\put( 8,7.5){\line(1,0){1}}
\put(7.5, 6){\line(0,1){1}}
\put(7.5, 4){\line(0,1){1}}
\put(7.5, 2){\line(0,1){1}}
\linethickness{0.6mm}
\put(2,7.5){\line(1,0){5.5}}
\put(7.5,7.5){\line(0,1){5.5}}
\put(7.5,7.5){\circle*{1.5}}
\end{picture} }
\put(0,21){
\parbox[t]{15cm}{
\begin{eqnarray}
\label{V4}
&\ \ =&\ \ \ \ \ 
\bar\chi_{1,1}(x+e_2)\ \bar\psi_{2,2}(x-e_1)\ \bar\psi_{1,2}(x-e_1)\ 
\bar\chi_{2,1}(x+e_2)\nonumber\\ 
&&\ \times\ \chi_{1,2}(x+e_2)\ \psi_{2,1}(x-e_1)\ \psi_{1,1}(x-e_1)\ 
\chi_{2,2}(x+e_2)/4
\end{eqnarray}  }}
\end{picture}

Vertex 6
\nopagebreak

\vspace{5mm}
\nopagebreak

\unitlength1.mm
\begin{picture}(150,20)
\put(0,5){
\unitlength1.mm
\begin{picture}(15,15)
\linethickness{0.15mm}
\put( 6,7.5){\line(1,0){1}}
\put( 4,7.5){\line(1,0){1}}
\put( 2,7.5){\line(1,0){1}}
\put(7.5, 6){\line(0,1){1}}
\put(7.5, 4){\line(0,1){1}}
\put(7.5, 2){\line(0,1){1}}
\linethickness{0.6mm}
\put(7.5,7.5){\line(1,0){5.5}}
\put(7.5,7.5){\line(0,1){5.5}}
\put(7.5,7.5){\circle*{1.5}}
\end{picture} }
\put(0,21){
\parbox[t]{15cm}{
\begin{eqnarray}
\label{V5}
&\ \ =&\ \ \ \ \ 
\bar\chi_{1,1}(x+e_2)\ \bar\psi_{2,1}(x+e_1)\ \bar\psi_{1,1}(x+e_1)\ 
\bar\chi_{2,1}(x+e_2)\nonumber\\ 
&&\ \times\ \chi_{1,2}(x+e_2)\ \psi_{2,2}(x+e_1)\ \psi_{1,2}(x+e_1)\ 
\chi_{2,2}(x+e_2)/4
\end{eqnarray}  }}
\end{picture}

Vertex 7
\nopagebreak

\vspace{5mm}
\nopagebreak

\unitlength1.mm
\begin{picture}(150,20)
\put(0,5){
\unitlength1.mm
\begin{picture}(15,15)
\linethickness{0.15mm}
\put(12,7.5){\line(1,0){1}}
\put(10,7.5){\line(1,0){1}}
\put( 8,7.5){\line(1,0){1}}
\put(7.5,12){\line(0,1){1}}
\put(7.5,10){\line(0,1){1}}
\put(7.5, 8){\line(0,1){1}}
\linethickness{0.6mm}
\put(2,7.5){\line(1,0){5.5}}
\put(7.5,2){\line(0,1){5.5}}
\put(7.5,7.5){\circle*{1.5}}
\end{picture} }
\put(0,21){
\parbox[t]{15cm}{
\begin{eqnarray}
\label{V7}
&\ \ =&\ \ \ \ \ 
\bar\chi_{1,2}(x-e_2)\ \bar\psi_{2,2}(x-e_1)\ \bar\psi_{1,2}(x-e_1)\ 
\bar\chi_{2,2}(x-e_2)\nonumber\\ 
&&\ \times\ \chi_{1,1}(x-e_2)\ \psi_{2,1}(x-e_1)\ \psi_{1,1}(x-e_1)\ 
\chi_{2,1}(x-e_2)/4
\end{eqnarray}  }}
\end{picture}

Vertex 8
\nopagebreak

\vspace{5mm}
\nopagebreak

\unitlength1.mm
\begin{picture}(150,25)
\put(0,10){
\unitlength1.mm
\begin{picture}(15,15)
\linethickness{0.15mm}
\put(12,7.5){\line(1,0){1}}
\put(10,7.5){\line(1,0){1}}
\put( 8,7.5){\line(1,0){1}}
\put(7.5,2){\line(0,1){11}}
\linethickness{0.6mm}
\put(2,7.5){\line(1,0){5.5}}
\put(7.5,7.5){\circle*{1.5}}
\end{picture} }
\put(0,26){
\parbox[t]{15cm}{
\begin{eqnarray}
\label{V18}
\hspace{10mm} =&-&\Big[\bar\chi_{1,1}(x+e_2)\ \bar\chi_{2,2}(x-e_2)\ + 
 \ \bar\chi_{1,2}(x-e_2)\ \bar\chi_{2,1}(x+e_2) \Big]\nonumber\\
&\times&\Big[\chi_{1,2}(x+e_2)\ \chi_{2,1}(x-e_2)\ + 
 \ \chi_{1,1}(x-e_2)\ \chi_{2,2}(x+e_2) \Big]\nonumber\\
&\times&\bar\psi_{1,2}(x-e_1)\ \bar\psi_{2,2}(x-e_1)\ 
 \psi_{1,1}(x-e_1)\ \psi_{2,1}(x-e_1) /4
\end{eqnarray}  }}
\end{picture}

Vertex 9
\nopagebreak

\vspace{5mm}
\nopagebreak

\unitlength1.mm
\begin{picture}(150,25)
\put(0,10){
\unitlength1.mm
\begin{picture}(15,15)
\linethickness{0.15mm}
\put( 6,7.5){\line(1,0){1}}
\put( 4,7.5){\line(1,0){1}}
\put( 2,7.5){\line(1,0){1}}
\put(7.5,2){\line(0,1){11}}
\linethickness{0.6mm}
\put(7.5,7.5){\line(1,0){5.5}}
\put(7.5,7.5){\circle*{1.5}}
\end{picture} }
\put(0,26){
\parbox[t]{15cm}{
\begin{eqnarray}
\label{V12}
\hspace{10mm} =&-& \Big[\bar\chi_{1,1}(x+e_2)\ \bar\chi_{2,2}(x-e_2)\ + 
 \ \bar\chi_{1,2}(x-e_2)\ \bar\chi_{2,1}(x+e_2) \Big]\nonumber\\
&\times&\Big[\chi_{1,2}(x+e_2)\ \chi_{2,1}(x-e_2)\ + 
 \ \chi_{1,1}(x-e_2)\ \chi_{2,2}(x+e_2) \Big]\nonumber\\
&\times&\bar\psi_{1,1}(x+e_1)\ \bar\psi_{2,1}(x+e_1)\ 
 \psi_{1,2}(x+e_1)\ \psi_{2,2}(x+e_1)/4
\end{eqnarray}  }}
\end{picture}

Vertex 10
\nopagebreak

\vspace{5mm}
\nopagebreak

\unitlength1.mm
\begin{picture}(150,25)
\put(0,10){
\unitlength1.mm
\begin{picture}(15,15)
\linethickness{0.15mm}
\put(2,7.5){\line(1,0){11}}
\put(7.5, 6){\line(0,1){1}}
\put(7.5, 4){\line(0,1){1}}
\put(7.5, 2){\line(0,1){1}}
\linethickness{0.6mm}
\put(7.5,7.5){\line(0,1){5.5}}
\put(7.5,7.5){\circle*{1.5}}
\end{picture} }
\put(0,26){
\parbox[t]{15cm}{
\begin{eqnarray}
\label{V9}
\hspace{10mm} =&-&  \Big[\bar\psi_{1,1}(x+e_1)\ \bar\psi_{2,2}(x-e_1)\ + 
  \ \bar\psi_{1,2}(x-e_1)\ \bar\psi_{2,1}(x+e_1) \Big]\nonumber\\
&\times& \Big[\psi_{1,2}(x+e_1)\ \psi_{2,1}(x-e_1)\ + 
  \ \psi_{1,1}(x-e_1)\ \psi_{2,2}(x+e_1) \Big]\nonumber\\
&\times& \bar\chi_{1,1}(x+e_2)\ \bar\chi_{2,1}(x+e_2)\ 
\chi_{1,2}(x+e_2)\ \chi_{2,2}(x+e_2)/4
\end{eqnarray}  }}
\end{picture}

Vertex 11
\nopagebreak

\vspace{5mm}
\nopagebreak

\unitlength1.mm
\begin{picture}(150,25)
\put(0,10){
\unitlength1.mm
\begin{picture}(15,15)
\linethickness{0.15mm}
\put(2,7.5){\line(1,0){11}}
\put(7.5,12){\line(0,1){1}}
\put(7.5,10){\line(0,1){1}}
\put(7.5, 8){\line(0,1){1}}
\linethickness{0.6mm}
\put(7.5,2){\line(0,1){5.5}}
\put(7.5,7.5){\circle*{1.5}}
\end{picture} }
\put(0,26){
\parbox[t]{15cm}{
\begin{eqnarray}
\label{V15}
\hspace{10mm} =&-&\Big[\bar\psi_{1,1}(x+e_1)\ \bar\psi_{2,2}(x-e_1)\ + 
  \ \bar\psi_{1,2}(x-e_1)\ \bar\psi_{2,1}(x+e_1) \Big]\nonumber\\
&\times&\Big[\psi_{1,2}(x+e_1)\ \psi_{2,1}(x-e_1)\ + 
  \ \psi_{1,1}(x-e_1)\ \psi_{2,2}(x+e_1) \Big]\nonumber\\
&\times&\bar\chi_{1,2}(x-e_2)\ \bar\chi_{2,2}(x-e_2)\ 
\chi_{1,1}(x-e_2)\ \chi_{2,1}(x-e_2) /4
\end{eqnarray}  }}
\end{picture}

Vertex 12
\nopagebreak

\vspace{5mm}
\nopagebreak

\unitlength1.mm
\begin{picture}(150,25)
\put(0,10){
\unitlength1.mm
\begin{picture}(15,15)
\linethickness{0.15mm}
\put(7.5,12){\line(0,1){1}}
\put(7.5,10){\line(0,1){1}}
\put(7.5, 8){\line(0,1){1}}
\put(7.5,7.5){\line(1,0){5.5}}
\put(7.5,2){\line(0,1){5.5}}
\linethickness{0.6mm}
\put(2,7.5){\line(1,0){5.5}}
\put(7.5,7.5){\circle*{1.5}}
\end{picture} }
\put(0,26){
\parbox[t]{15cm}{
\begin{eqnarray}
\label{V17}
\hspace{10mm} =&\ \ &\Big[\bar\chi_{1,2}(x-e_2)\ \bar\psi_{2,1}(x+e_1)\  + 
     \ \bar\psi_{1,1}(x+e_1)\ \bar\chi_{2,2}(x-e_2) \Big]\nonumber\\
&\times&\Big[\chi_{1,1}(x-e_2)\ \psi_{2,2}(x+e_1)\  + 
     \ \psi_{1,2}(x+e_1)\ \chi_{2,1}(x-e_2) \Big]\nonumber\\
&\times&\bar\psi_{1,2}(x-e_1)\ \bar\psi_{2,2}(x-e_1)\ 
\psi_{1,1}(x-e_1)\ \psi_{2,1}(x-e_1)/2
\end{eqnarray}  }}
\end{picture}

Vertex 13
\nopagebreak

\vspace{5mm}
\nopagebreak

\unitlength1.mm
\begin{picture}(150,25)
\put(0,10){
\unitlength1.mm
\begin{picture}(15,15)
\linethickness{0.15mm}
\put( 6,7.5){\line(1,0){1}}
\put( 4,7.5){\line(1,0){1}}
\put( 2,7.5){\line(1,0){1}}
\put(7.5,2){\line(0,1){5.5}}
\put(7.5,7.5){\line(1,0){5.5}}
\linethickness{0.6mm}
\put(7.5,7.5){\line(0,1){5.5}}
\put(7.5,7.5){\circle*{1.5}}
\end{picture} }
\put(0,26){
\parbox[t]{15cm}{
\begin{eqnarray}
\label{V14}
\hspace{10mm} =&\ \ &\Big[\bar\chi_{1,2}(x-e_2)\ \bar\psi_{2,1}(x+e_1) \ + 
     \ \bar\psi_{1,1}(x+e_1)\ \bar\chi_{2,2}(x-e_2) \Big]\nonumber\\
&\times&\Big[\chi_{1,1}(x-e_2)\ \psi_{2,2}(x+e_1) \ + 
     \ \psi_{1,2}(x+e_1)\ \chi_{2,1}(x-e_2) \Big]\nonumber\\
&\times&\bar\chi_{1,1}(x+e_2)\ \bar\chi_{2,1}(x+e_2)\ 
\chi_{1,2}(x+e_2)\ \chi_{2,2}(x+e_2)/2
\end{eqnarray}  }}
\end{picture}

Vertex 14
\nopagebreak

\vspace{5mm}
\nopagebreak

\unitlength1.mm
\begin{picture}(150,25)
\put(0,10){
\unitlength1.mm
\begin{picture}(15,15)
\linethickness{0.15mm}
\put(7.5, 6){\line(0,1){1}}
\put(7.5, 4){\line(0,1){1}}
\put(7.5, 2){\line(0,1){1}}
\put(2,7.5){\line(1,0){5.5}}
\put(7.5,7.5){\line(0,1){5.5}}
\linethickness{0.6mm}
\put(7.5,7.5){\line(1,0){5.5}}
\put(7.5,7.5){\circle*{1.5}}
\end{picture} }
\put(0,26){
\parbox[t]{15cm}{
\begin{eqnarray}
\label{V10}
\hspace{10mm} =&\ \ & \Big[\bar\chi_{1,1}(x+e_2)\ \bar\psi_{2,2}(x-e_1)\  + 
      \ \bar\psi_{1,2}(x-e_1)\ \bar\chi_{2,1}(x+e_2)\Big]\nonumber\\ 
&\times& \Big[\chi_{1,2}(x+e_2)\ \psi_{2,1}(x-e_1)\  + 
      \ \psi_{1,1}(x-e_1)\ \chi_{2,2}(x+e_2) \Big]\nonumber\\
&\times&\bar\psi_{1,1}(x+e_1)\ \bar\psi_{2,1}(x+e_1)\ 
\psi_{1,2}(x+e_1)\ \psi_{2,2}(x+e_1)/2
\end{eqnarray}  }}
\end{picture}

Vertex 15
\nopagebreak

\vspace{5mm}
\nopagebreak

\unitlength1.mm
\begin{picture}(150,25)
\put(0,10){
\unitlength1.mm
\begin{picture}(15,15)
\linethickness{0.15mm}
\put(12,7.5){\line(1,0){1}}
\put(10,7.5){\line(1,0){1}}
\put( 8,7.5){\line(1,0){1}}
\put(7.5,7.5){\line(0,1){5.5}}
\put(2,7.5){\line(1,0){5.5}}
\linethickness{0.6mm}
\put(7.5,2){\line(0,1){5.5}}
\put(7.5,7.5){\circle*{1.5}}
\end{picture} }
\put(0,26){
\parbox[t]{15cm}{
\begin{eqnarray}
\label{V19}
\hspace{10mm} =&\ \ & \Big[\bar\chi_{1,1}(x+e_2)\ \bar\psi_{2,2}(x-e_1) \ + 
      \ \bar\psi_{1,2}(x-e_1)\ \bar\chi_{2,1}(x+e_2) \Big]\nonumber\\
&\times& \Big[\chi_{1,2}(x+e_2)\ \psi_{2,1}(x-e_1) \ + 
      \ \psi_{1,1}(x-e_1)\ \chi_{2,2}(x+e_2) \Big]\nonumber\\
&\times&\bar\chi_{1,2}(x-e_2)\ \bar\chi_{2,2}(x-e_2)\ 
\chi_{1,1}(x-e_2)\ \chi_{2,1}(x-e_2)/2
\end{eqnarray}  }}
\end{picture}

Vertex 16
\nopagebreak

\vspace{5mm}
\nopagebreak

\unitlength1.mm
\begin{picture}(150,25)
\put(0,10){
\unitlength1.mm
\begin{picture}(15,15)
\linethickness{0.15mm}
\put(7.5, 6){\line(0,1){1}}
\put(7.5, 4){\line(0,1){1}}
\put(7.5, 2){\line(0,1){1}}
\put(7.5,7.5){\line(1,0){5.5}}
\put(7.5,7.5){\line(0,1){5.5}}
\linethickness{0.6mm}
\put(2,7.5){\line(1,0){5.5}}
\put(7.5,7.5){\circle*{1.5}}
\end{picture} }
\put(0,26){
\parbox[t]{15cm}{
\begin{eqnarray}
\label{V11}
\hspace{10mm} =&-&\Big[\bar\psi_{1,1}(x+e_1)\ \bar\chi_{2,1}(x+e_2)\ + 
 \ \bar\chi_{1,1}(x+e_2)\ \bar\psi_{2,1}(x+e_1) \Big]\nonumber\\
&\times&\Big[\psi_{1,2}(x+e_1)\ \chi_{2,2}(x+e_2)\ + 
 \ \chi_{1,2}(x+e_2)\ \psi_{2,2}(x+e_1) \Big]\nonumber\\
&\times&\bar\psi_{1,2}(x-e_1)\ \bar\psi_{2,2}(x-e_1)\ 
 \psi_{1,1}(x-e_1)\ \psi_{2,1}(x-e_1)/2
\end{eqnarray}  }}
\end{picture}

Vertex 17
\nopagebreak

\vspace{5mm}
\nopagebreak

\unitlength1.mm
\begin{picture}(150,25)
\put(0,10){
\unitlength1.mm
\begin{picture}(15,15)
\linethickness{0.15mm}
\put( 6,7.5){\line(1,0){1}}
\put( 4,7.5){\line(1,0){1}}
\put( 2,7.5){\line(1,0){1}}
\put(7.5,7.5){\line(0,1){5.5}}
\put(7.5,7.5){\line(1,0){5.5}}
\linethickness{0.6mm}
\put(7.5,2){\line(0,1){5.5}}
\put(7.5,7.5){\circle*{1.5}}
\end{picture} }
\put(0,26){
\parbox[t]{15cm}{
\begin{eqnarray}
\label{V13}
\hspace{10mm} =&-& \Big[\ \bar\psi_{1,1}(x+e_1)\ \bar\chi_{2,1}(x+e_2)+ 
 \ \bar\chi_{1,1}(x+e_2)\ \bar\psi_{2,1}(x+e_1) \Big]\nonumber\\
&\times& \Big[\ \psi_{1,2}(x+e_1)\ \chi_{2,2}(x+e_2)+ 
 \ \chi_{1,2}(x+e_2)\ \psi_{2,2}(x+e_1) \Big]\nonumber\\
&\times&\bar\chi_{1,2}(x-e_2)\ \bar\chi_{2,2}(x-e_2)\ 
 \chi_{1,1}(x-e_2)\ \chi_{2,1}(x-e_2)/2
\end{eqnarray}  }}
\end{picture}

Vertex 18
\nopagebreak

\vspace{5mm}
\nopagebreak

\unitlength1.mm
\begin{picture}(150,25)
\put(0,10){
\unitlength1.mm
\begin{picture}(15,15)
\linethickness{0.15mm}
\put(12,7.5){\line(1,0){1}}
\put(10,7.5){\line(1,0){1}}
\put( 8,7.5){\line(1,0){1}}
\put(7.5,2){\line(0,1){5.5}}
\put(2,7.5){\line(1,0){5.5}}
\linethickness{0.6mm}
\put(7.5,7.5){\line(0,1){5.5}}
\put(7.5,7.5){\circle*{1.5}}
\end{picture} }
\put(0,26){
\parbox[t]{15cm}{
\begin{eqnarray}
\label{V20}
\hspace{10mm} =&-&\Big[\bar\psi_{1,2}(x-e_1)\ \bar\chi_{2,2}(x-e_2)\ + 
\ \bar\chi_{1,2}(x-e_2)\ \bar\psi_{2,2}(x-e_1) \Big]\nonumber\\
&\times&\Big[\psi_{1,1}(x-e_1)\ \chi_{2,1}(x-e_2)\ + 
\ \chi_{1,1}(x-e_2)\ \psi_{2,1}(x-e_1) \Big]\nonumber\\
&\times&\bar\chi_{1,1}(x+e_2)\ \bar\chi_{2,1}(x+e_2)\ 
\chi_{1,2}(x+e_2)\ \chi_{2,2}(x+e_2)/2
\end{eqnarray}  }}
\end{picture}

Vertex 19
\nopagebreak

\vspace{5mm}
\nopagebreak

\unitlength1.mm
\begin{picture}(150,25)
\put(0,10){
\unitlength1.mm
\begin{picture}(15,15)
\linethickness{0.15mm}
\put(7.5,12){\line(0,1){1}}
\put(7.5,10){\line(0,1){1}}
\put(7.5, 8){\line(0,1){1}}
\put(2,7.5){\line(1,0){5.5}}
\put(7.5,2){\line(0,1){5.5}}
\linethickness{0.6mm}
\put(7.5,7.5){\line(1,0){5.5}}
\put(7.5,7.5){\circle*{1.5}}
\end{picture} }
\put(0,26){
\parbox[t]{15cm}{
\begin{eqnarray}
\label{V16}
\hspace{10mm} =&-& \Big[\bar\psi_{1,2}(x-e_1)\ \bar\chi_{2,2}(x-e_2)\ + 
 \ \bar\chi_{1,2}(x-e_2)\ \bar\psi_{2,2}(x-e_1) \Big]\nonumber\\
&\times& \Big[\psi_{1,1}(x-e_1)\ \chi_{2,1}(x-e_2)\ + 
 \ \chi_{1,1}(x-e_2)\ \psi_{2,1}(x-e_1) \Big]\nonumber\\
&\times&  \bar\psi_{1,1}(x+e_1)\ \bar\psi_{2,1}(x+e_1)\ 
\psi_{1,2}(x+e_1)\ \psi_{2,2}(x+e_1)/2
\end{eqnarray}  }}
\end{picture}

Vertex 20
\nopagebreak

\vspace{5mm}
\nopagebreak

\unitlength1.mm
\begin{picture}(150,90)
\put(0,75){
\unitlength1.mm
\begin{picture}(15,15)
\linethickness{0.15mm}
\put(2,7.5){\line(1,0){11}}
\put(7.5,2){\line(0,1){11}}
\put(7.5,7.5){\circle*{1.5}}
\end{picture} }
\put(0,91){
\parbox[t]{15cm}{
\begin{eqnarray}
\label{V8}
&&\hspace{-2mm}=\nonumber\\[8mm]
\Bigg[&& \Big[\bar\chi_{1,1}(x+e_2)\ \bar\psi_{2,2}(x-e_1)\  + 
      \ \bar\psi_{1,2}(x-e_1)\ \bar\chi_{2,1}(x+e_2) \Big]\nonumber\\
&\times&\Big[\bar\chi_{1,2}(x-e_2)\ \bar\psi_{2,1}(x+e_1) \ + 
     \ \bar\psi_{1,1}(x+e_1)\ \bar\chi_{2,2}(x-e_2) \Big]\nonumber\\
&-& \Big[\bar\psi_{1,2}(x-e_1)\ \bar\chi_{2,2}(x-e_2)\ + 
 \ \bar\chi_{1,2}(x-e_2)\ \bar\psi_{2,2}(x-e_1) \Big]\nonumber\\
&\times& \Big[\bar\psi_{1,1}(x+e_1)\ \bar\chi_{2,1}(x+e_2)\  + 
 \ \bar\chi_{1,1}(x+e_2)\ \bar\psi_{2,1}(x+e_1) \Big]
\hspace{0.5cm} \Bigg]\nonumber\\
\times\ \Bigg[&& \Big[\chi_{1,2}(x+e_2)\ \psi_{2,1}(x-e_1)\  + 
      \ \psi_{1,1}(x-e_1)\ \chi_{2,2}(x+e_2) \Big]\nonumber\\
&\times&\Big[\chi_{1,1}(x-e_2)\ \psi_{2,2}(x+e_1) \ + 
     \ \psi_{1,2}(x+e_1)\ \chi_{2,1}(x-e_2) \Big]\nonumber\\
&-& \Big[\psi_{1,1}(x-e_1)\ \chi_{2,1}(x-e_2)\ + 
 \ \chi_{1,1}(x-e_2)\ \psi_{2,1}(x-e_1) \Big]\nonumber\\
&\times&\Big[\psi_{1,2}(x+e_1)\ \chi_{2,2}(x+e_2)\  + 
 \ \chi_{1,2}(x+e_2)\ \psi_{2,2}(x+e_1) \Big] \hspace{0.5cm} \Bigg]
\Bigg/ 4
\end{eqnarray}  }}
\end{picture}

\vspace{3mm}

Remarkably, as one recognizes from the above equations the remaining 
field combinations on the odd sublattice factorize for each vertex 
into exactly identical contributions from the $\bar\psi$, $\bar\chi$
and the $\psi$, $\chi$ subsystems. To see this use the map 

\begin{eqnarray}
\label{B8}
&\bar\psi_{i,1}(x)&\longleftrightarrow 
- -\ \psi_{i,2}(x)\ \ \ ,\nonumber\\[0.3cm]
&\bar\psi_{i,2}(x)&\longleftrightarrow 
\ \ \ \psi_{i,1}(x)
\end{eqnarray}

entailing

\begin{eqnarray}
\label{B9}
&\bar\chi_{i,1}(x)&\longleftrightarrow
\ \ \ \chi_{i,2}(x)\ \ \ ,\nonumber\\[0.3cm] 
&\bar\chi_{i,2}(x)&\longleftrightarrow
- -\ \chi_{i,1}(x)\ \ \ .
\end{eqnarray}

The above mentioned factorization property for each vertex comes as a 
surprise because it could hardly have been guessed prior to the explicit 
calculation. The vertex factorization property immediately leads to 
the conclusion that any individual contribution (of any vertex cluster) 
to the partition function $Z_\Lambda$ is positive. The minus 
sign in front of the r.h.s.\ of eqs.\ (\ref{V18})-(\ref{V15}), 
(\ref{V11})-(\ref{V16}) at first glance 
disturbing this property is properly
taken care of by the map (\ref{B8}), (\ref{B9}).
It should be mentioned that a similar vertex factorization 
property holds in the one-flavour Schwinger model \cite{salm}, there
however being fulfilled rather trivially. One might wonder, 
whether a vertex factorization property also holds for the strong
coupling Schwinger
model with a larger number of fermion species than two, 
perhaps being a rather general structure.\\

\parindent1.5em

The vertex factorization property means that we can now restrict
ourselves to the discussion of the $\psi$, $\chi$ subsystem. 
Each vertex cluster contributing to the partition function $Z_\Lambda$
adds a term to the partition sum equal to the square of the 
weight of the corresponding cluster in the $\psi$, $\chi$ subsystem.
We consequently need to know weights in the $\psi$, $\chi$ subsystem
up to minus signs only.
In order to achieve further understanding it is useful to assign the remaining 
$\psi$, $\chi$ fields on the odd sublattice $\Lambda_o$ certain  
graphical symbols as follows (The full black dot denotes a 
point $x$ on the even sublattice while the hollow circle stands 
for the point on the odd sublattice which the argument
of the fields relates to.).\\

\vspace{3mm}

\parindent0.em

\unitlength1.mm
\begin{picture}(150,7)
\put(0,5){
\unitlength1.mm
\begin{picture}(8.25,2)
\linethickness{0.15mm}
\put(2,1){\vector(1,0){3.75}}
\put(7.5,1){\line(-1,0){1.75}}
\put(1,1){\circle{2}}
\put(7.5,1){\circle*{1.5}}
\end{picture}  }
\put(70,5){
\unitlength1.mm
\begin{picture}(8.25,2)
\linethickness{0.15mm}
\put(7.5,1){\vector(-1,0){3.75}}
\put(2,1){\line(1,0){1.75}}
\put(1,1){\circle{2}}
\put(7.5,1){\circle*{1.5}}
\end{picture}   }
\put(0,6){
\parbox[t]{15cm}{
\parbox{6cm}{
\begin{eqnarray}
\label{F1}
=&\psi_{1,1}(x-e_1)&
\end{eqnarray}  }
\parbox{7mm}{\ }
\parbox{6cm}{
\begin{eqnarray}
\label{F2}
=&\psi_{2,1}(x-e_1)&
\end{eqnarray}  }  }}
\end{picture} 

\vspace{6mm}

\unitlength1.mm
\begin{picture}(150,7)
\put(0,5){
\unitlength1.mm
\begin{picture}(8.25,2)
\linethickness{0.15mm}
\put(0.75,1){\vector(1,0){3.75}}
\put(6.25,1){\line(-1,0){1.75}}
\put(7.25,1){\circle{2}}
\put(0.75,1){\circle*{1.5}}
\end{picture}   }
\put(70,5){
\unitlength1.mm
\begin{picture}(8.25,2)
\linethickness{0.15mm}
\put(6.25,1){\vector(-1,0){3.75}}
\put(0.75,1){\line(1,0){1.75}}
\put(7.25,1){\circle{2}}
\put(0.75,1){\circle*{1.5}}
\end{picture}  }
\put(0,6){
\parbox[t]{15cm}{
\parbox{6cm}{
\begin{eqnarray}
\label{F3}
=&\psi_{2,2}(x+e_1)&
\end{eqnarray}  }
\parbox{7mm}{\ }
\parbox{6cm}{
\begin{eqnarray}
\label{F4}
=&\psi_{1,2}(x+e_1)&
\end{eqnarray}  } }}
\end{picture}

\vspace{3mm}

\unitlength1.mm
\begin{picture}(150,8)
\put(3,0){
\unitlength1.mm
\begin{picture}(2,8.25)
\linethickness{0.15mm}
\put(1,2){\vector(0,1){3.75}}
\put(1,7.5){\line(0,-1){1.75}}
\put(1,1){\circle{2}}
\put(1,7.5){\circle*{1.5}}
\end{picture}  }
\put(73,0){
\unitlength1.mm
\begin{picture}(2,8.25)
\linethickness{0.15mm}
\put(1,7.5){\vector(0,-1){3.75}}
\put(1,2){\line(0,1){1.75}}
\put(1,1){\circle{2}}
\put(1,7.5){\circle*{1.5}}
\end{picture}    }
\put(0,4){
\parbox[t]{15cm}{
\parbox{6cm}{
\begin{eqnarray}
\label{F5}
=&\chi_{1,1}(x-e_2)&
\end{eqnarray}  }
\parbox{7mm}{\ }
\parbox{6cm}{
\begin{eqnarray}
\label{F6}
=&\chi_{2,1}(x-e_2)&
\end{eqnarray}  } }}
\end{picture} 

\vspace{4mm}

\unitlength1.mm
\begin{picture}(150,8)
\put(3,0){
\unitlength1.mm
\begin{picture}(2,8.25)
\linethickness{0.15mm}
\put(1,0.75){\vector(0,1){3.75}}
\put(1,6.25){\line(0,-1){1.75}}
\put(1,7.25){\circle{2}}
\put(1,0.75){\circle*{1.5}}
\end{picture}  }
\put(73,0){
\unitlength1.mm
\begin{picture}(2,8.25)
\linethickness{0.15mm}
\put(1,6.25){\vector(0,-1){3.75}}
\put(1,0.75){\line(0,1){1.75}}
\put(1,7.25){\circle{2}}
\put(1,0.75){\circle*{1.5}}
\end{picture}    }
\put(0,4){
\parbox[t]{15cm}{
\parbox{6cm}{
\begin{eqnarray}
\label{F7}
=&\chi_{2,2}(x+e_2)&
\end{eqnarray}  }
\parbox{7mm}{\ }
\parbox{6cm}{
\begin{eqnarray}
\label{F8}
=&\chi_{1,2}(x+e_2)&
\end{eqnarray}  }  }}
\end{picture} 

\vspace{7mm}

As a rule, an arrow flowing out of a point of the odd sublattice
(hollow circles) denotes a field with flavour index 1 while 
an arrow flowing into a point on the odd sublattice relates 
to a field with flavour index 2. From this it is already
clear that only those products of (four) Grassmann fields 
at any point of the odd sublattice allow non-vanishing 
contributions to the partition function $Z_\Lambda$ 
which have two incoming and two outgoing arrows in their
graphical symbols. We will refer to this fact as the 
vertex arrow rule.
The symbols for the fields on the odd sublattice can now be 
used in a natural way to construct further graphical building blocks.\\[5mm]

\unitlength1.mm
\begin{picture}(150,30)
\put(6.5,15){
\unitlength1.mm
\begin{picture}(2,15)
\linethickness{0.15mm}
\put(1,2){\line(0,1){11}}
\put(1,14){\circle{2}}
\put(1,7.5){\circle*{1.5}}
\put(1,1){\circle{2}}
\end{picture}  }
\put(51.5,15){
\unitlength1.mm
\begin{picture}(2,15)
\linethickness{0.15mm}
\put(1,2){\vector(0,1){3.75}}
\put(1,7.5){\line(0,-1){1.75}}
\put(1,7.5){\vector(0,1){3.75}}
\put(1,13){\line(0,-1){1.75}}
\put(1,1){\circle{2}}
\put(1,7.5){\circle*{1.5}}
\put(1,14){\circle{2}}
\end{picture}  }
\put(106.5,15){
\unitlength1.mm
\begin{picture}(2,15)
\linethickness{0.15mm}
\put(1,13){\vector(0,-1){3.75}}
\put(1,7.5){\line(0,1){1.75}}
\put(1,7.5){\vector(0,-1){3.75}}
\put(1,2){\line(0,1){1.75}}
\put(1,14){\circle{2}}
\put(1,7.5){\circle*{1.5}}
\put(1,1){\circle{2}}
\end{picture}  }
\put(0,30.75){
\parbox[t]{15cm}{
\begin{eqnarray}
\label{S2}
=&&\hspace{4.95cm}+\nonumber\\[11.5mm]
=&&\ \ \chi_{1,1}(x-e_2)\ \chi_{2,2}(x+e_2)\ \ +\ \ 
\chi_{1,2}(x+e_2)\ \chi_{2,1}(x-e_2)
\end{eqnarray} }}
\end{picture} 

\vspace{7mm}

\unitlength1.mm
\begin{picture}(150,20)
\put(0,18){
\unitlength1.mm
\begin{picture}(15,2)
\linethickness{0.15mm}
\put(2,1){\line(1,0){11}}
\put(1,1){\circle{2}}
\put(7.5,1){\circle*{1.5}}
\put(14,1){\circle{2}}
\end{picture} }
\put(45,18){
\unitlength1.mm
\begin{picture}(15,2)
\linethickness{0.15mm}
\put(2,1){\vector(1,0){3.75}}
\put(7.5,1){\line(-1,0){1.75}}
\put(7.5,1){\vector(1,0){3.75}}
\put(13,1){\line(-1,0){1.75}}
\put(1,1){\circle{2}}
\put(7.5,1){\circle*{1.5}}
\put(14,1){\circle{2}}
\end{picture}  }
\put(100,18){
\unitlength1.mm
\begin{picture}(15,2)
\linethickness{0.15mm}
\put(13,1){\vector(-1,0){3.75}}
\put(7.5,1){\line(1,0){1.75}}
\put(7.5,1){\vector(-1,0){3.75}}
\put(2,1){\line(1,0){1.75}}
\put(14,1){\circle{2}}
\put(7.5,1){\circle*{1.5}}
\put(1,1){\circle{2}}
\end{picture}   }
\put(0,27.5){
\parbox[t]{15cm}{
\begin{eqnarray}
\label{S1}
=&&\hspace{4.97cm}+\nonumber\\[0.5cm]
=&&\ \ \psi_{1,1}(x-e_1)\ \psi_{2,2}(x+e_1)\ \ +\ \ 
\psi_{1,2}(x+e_1)\ \psi_{2,1}(x-e_1)
\end{eqnarray}  }}
\end{picture} 

\vspace{7mm}

\unitlength1.mm
\begin{picture}(150,25)
\put(6.5,16.5){
\unitlength1.mm
\begin{picture}(8.5,8.5)
\linethickness{0.15mm}
\put(1,7.5){\line(1,0){5.5}}
\put(1,7.5){\line(0,-1){5.5}}
\put(7.5,7.5){\circle{2}}
\put(1,7.5){\circle*{1.5}}
\put(1,1){\circle{2}}
\end{picture}  }
\put(51.5,16.5){
\unitlength1.mm
\begin{picture}(8.5,8.5)
\linethickness{0.15mm}
\put(1,2){\vector(0,1){3.75}}
\put(1,7.5){\line(0,-1){1.75}}
\put(1,7.5){\vector(1,0){3.75}}
\put(6.5,7.5){\line(-1,0){1.75}}
\put(1,1){\circle{2}}
\put(1,7.5){\circle*{1.5}}
\put(7.5,7.5){\circle{2}}
\end{picture}  }
\put(106.5,16.5){
\unitlength1.mm
\begin{picture}(8.5,8.5)
\linethickness{0.15mm}
\put(6.5,7.5){\vector(-1,0){3.75}}
\put(1,7.5){\line(1,0){1.75}}
\put(1,7.5){\vector(0,-1){3.75}}
\put(1,2){\line(0,1){1.75}}
\put(7.5,7.5){\circle{2}}
\put(1,7.5){\circle*{1.5}}
\put(1,1){\circle{2}}
\end{picture} }
\put(0,32.5){
\parbox[t]{15cm}{
\begin{eqnarray}
\label{S5}
=&&\hspace{4.95cm}+\nonumber\\[11.5mm]
=&&\ \ \chi_{1,1}(x-e_2)\ \psi_{2,2}(x+e_1)\ \ +\ \ 
\psi_{1,2}(x+e_1)\ \chi_{2,1}(x-e_2)
\end{eqnarray}  }}
\end{picture} 

\vspace{7mm}

\unitlength1.mm
\begin{picture}(150,25)
\put(0,16.5){
\unitlength1.mm
\begin{picture}(8.5,8.5)
\linethickness{0.15mm}
\put(7.5,1){\line(-1,0){5.5}}
\put(7.5,1){\line(0,1){5.5}}
\put(7.5,7.5){\circle{2}}
\put(7.5,1){\circle*{1.5}}
\put(1,1){\circle{2}}
\end{picture}  }
\put(45,16.5){
\unitlength1.mm
\begin{picture}(8.5,8.5)
\linethickness{0.15mm}
\put(2,1){\vector(1,0){3.75}}
\put(7.5,1){\line(-1,0){1.75}}
\put(7.5,1){\vector(0,1){3.75}}
\put(7.5,6.5){\line(0,-1){1.75}}
\put(1,1){\circle{2}}
\put(7.5,1){\circle*{1.5}}
\put(7.5,7.5){\circle{2}}
\end{picture}  }
\put(100,16.5){
\unitlength1.mm
\begin{picture}(8.5,8.5)
\linethickness{0.15mm}
\put(7.5,6.5){\vector(0,-1){3.75}}
\put(7.5,1){\line(0,1){1.75}}
\put(7.5,1){\vector(-1,0){3.75}}
\put(2,1){\line(1,0){1.75}}
\put(7.5,7.5){\circle{2}}
\put(7.5,1){\circle*{1.5}}
\put(1,1){\circle{2}}
\end{picture} }
\put(0,26){
\parbox[t]{15cm}{
\begin{eqnarray}
\label{S4}
=&&\hspace{4.95cm}+\nonumber\\[5mm]
=&&\ \ \psi_{1,1}(x-e_1)\ \chi_{2,2}(x+e_2)\ \ +\ \ 
\chi_{1,2}(x+e_2)\ \psi_{2,1}(x-e_1)
\end{eqnarray}  }}
\end{picture} 

\vspace{7mm}

\unitlength1.mm
\begin{picture}(150,25)
\put(6.5,16.5){
\unitlength1.mm
\begin{picture}(8.5,8.5)
\linethickness{0.15mm}
\put(1,1){\line(0,1){5.5}}
\put(1,1){\line(1,0){5.5}}
\put(1,7.5){\circle{2}}
\put(1,1){\circle*{1.5}}
\put(7.5,1){\circle{2}}
\end{picture}  }
\put(51.5,16.5){
\unitlength1.mm
\begin{picture}(8.5,8.5)
\linethickness{0.15mm}
\put(1,6.5){\vector(0,-1){3.75}}
\put(1,1){\line(0,1){1.75}}
\put(1,1){\vector(1,0){3.75}}
\put(6.5,1){\line(-1,0){1.75}}
\put(1,7.5){\circle{2}}
\put(1,1){\circle*{1.5}}
\put(7.5,1){\circle{2}}
\end{picture}  }
\put(106.5,16.5){
\unitlength1.mm
\begin{picture}(8.5,8.5)
\linethickness{0.15mm}
\put(6.5,1){\vector(-1,0){3.75}}
\put(1,1){\line(1,0){1.75}}
\put(1,1){\vector(0,1){3.75}}
\put(1,6.5){\line(0,-1){1.75}}
\put(7.5,1){\circle{2}}
\put(1,1){\circle*{1.5}}
\put(1,7.5){\circle{2}}
\end{picture}  }
\put(0,26){
\parbox[t]{15cm}{
\begin{eqnarray}
\label{S3}
=&&\hspace{4.95cm}+\nonumber\\[5mm]
=&&\ \ \chi_{1,2}(x+e_2)\ \psi_{2,2}(x+e_1)\ \ +\ \ 
\psi_{1,2}(x+e_1)\ \chi_{2,2}(x+e_2)
\end{eqnarray}  }}
\end{picture} 

\vspace{7mm}

\unitlength1.mm
\begin{picture}(150,25)
\put(0,16.5){
\unitlength1.mm
\begin{picture}(8.5,8.5)
\linethickness{0.15mm}
\put(7.5,7.5){\line(0,-1){5.5}}
\put(7.5,7.5){\line(-1,0){5.5}}
\put(1,7.5){\circle{2}}
\put(7.5,7.5){\circle*{1.5}}
\put(7.5,1){\circle{2}}
\end{picture}  }
\put(45,16.5){
\unitlength1.mm
\begin{picture}(8.5,8.5)
\linethickness{0.15mm}
\put(2,7.5){\vector(1,0){3.75}}
\put(7.5,7.5){\line(-1,0){1.75}}
\put(7.5,7.5){\vector(0,-1){3.75}}
\put(7.5,2){\line(0,1){1.75}}
\put(1,7.5){\circle{2}}
\put(7.5,7.5){\circle*{1.5}}
\put(7.5,1){\circle{2}}
\end{picture}  }
\put(100,16.5){
\unitlength1.mm
\begin{picture}(2,8)
\linethickness{0.15mm}
\put(7.5,2){\vector(0,1){3.75}}
\put(7.5,7.5){\line(0,-1){1.75}}
\put(7.5,7.5){\vector(-1,0){3.75}}
\put(2,7.5){\line(1,0){1.75}}
\put(1,7.5){\circle{2}}
\put(7.5,7.5){\circle*{1.5}}
\put(7.5,1){\circle{2}}
\end{picture}  }
\put(0,32.5){
\parbox[t]{15cm}{
\begin{eqnarray}
\label{S6}
=&&\hspace{4.95cm}+\nonumber\\[11.5mm]
=&&\ \ \psi_{1,1}(x-e_1)\ \chi_{2,1}(x-e_2)\ \ +\ \ 
\chi_{1,1}(x-e_2)\ \psi_{2,1}(x-e_1)
\end{eqnarray}  }}
\end{picture}

\vspace{7mm}

\unitlength1.mm
\begin{picture}(150,8.25)
\put(3.125,0){
\unitlength1.mm
\begin{picture}(2,8.25)
\linethickness{0.6mm}
\put(1,2){\line(0,1){5.5}}
\put(1,1){\circle{2}}
\put(1,7.5){\circle*{1.5}}
\end{picture}  }
\put(36.625,0){
\unitlength1.mm
\begin{picture}(2,8.25)
\linethickness{0.15mm}
\put(1,2){\vector(0,1){3.75}}
\put(1,7.5){\line(0,-1){1.75}}
\put(1,1){\circle{2}}
\put(1,7.5){\circle*{1.5}}
\end{picture}  }
\put(54.625,0){
\unitlength1.mm
\begin{picture}(2,8.25)
\linethickness{0.15mm}
\put(1,7.5){\vector(0,-1){3.75}}
\put(1,2){\line(0,1){1.75}}
\put(1,1){\circle{2}}
\put(1,7.5){\circle*{1.5}}
\end{picture}    }
\put(0,12.5){
\parbox[t]{15cm}{
\begin{eqnarray}
\label{S9}
=\hspace{2cm}\times\hspace{1.5cm}&=&\chi_{1,1}(x-e_2)\ \chi_{2,1}(x-e_2)
\hspace{1.4cm}
\end{eqnarray}  }}
\end{picture} 

\vspace{3mm}

\unitlength1.mm
\begin{picture}(150,8.25)
\put(3.125,0){
\unitlength1.mm
\begin{picture}(2,8.25)
\linethickness{0.6mm}
\put(1,0.75){\line(0,1){5.5}}
\put(1,7.25){\circle{2}}
\put(1,0.75){\circle*{1.5}}
\end{picture}  }
\put(36.525,0){
\unitlength1.mm
\begin{picture}(2,8)
\linethickness{0.15mm}
\put(1,0.5){\vector(0,1){3.75}}
\put(1,6){\line(0,-1){1.75}}
\put(1,7){\circle{2}}
\put(1,0.5){\circle*{1.5}}
\end{picture}  }
\put(54.525,0){
\unitlength1.mm
\begin{picture}(2,8.25)
\linethickness{0.15mm}
\put(1,6){\vector(0,-1){3.75}}
\put(1,0.5){\line(0,1){1.75}}
\put(1,7){\circle{2}}
\put(1,0.5){\circle*{1.5}}
\end{picture}    }
\put(0,12.5){
\parbox[t]{15cm}{
\begin{eqnarray}
\label{S10}
=\hspace{2cm}\times\hspace{1.5cm}&=&\chi_{1,2}(x+e_2)\ \chi_{2,2}(x+e_2)
\hspace{1.4cm}
\end{eqnarray}  }}
\end{picture} 

\vspace{3mm}

\unitlength1.mm
\begin{picture}(150,8.5)
\put(0,3){
\unitlength1.mm
\begin{picture}(8.25,2)
\linethickness{0.6mm}
\put(2,1){\line(1,0){5.5}}
\put(1,1){\circle{2}}
\put(7.5,1){\circle*{1.5}}
\end{picture}  }
\put(33.5,3){
\unitlength1.mm
\begin{picture}(8.25,2)
\linethickness{0.15mm}
\put(2,1){\vector(1,0){3.75}}
\put(7.5,1){\line(-1,0){1.75}}
\put(1,1){\circle{2}}
\put(7.5,1){\circle*{1.5}}
\end{picture}  }
\put(51.5,3){
\unitlength1.mm
\begin{picture}(8.25,2)
\linethickness{0.15mm}
\put(7.5,1){\vector(-1,0){3.75}}
\put(2,1){\line(1,0){1.75}}
\put(1,1){\circle{2}}
\put(7.5,1){\circle*{1.5}}
\end{picture}   }
\put(0,12.5){
\parbox[t]{15cm}{
\begin{eqnarray}
\label{S7}
=\hspace{2cm}\times\hspace{1.5cm}&=&\psi_{1,1}(x-e_1)\ \psi_{2,1}(x-e_1)
\hspace{1.4cm}
\end{eqnarray}  }}
\end{picture} 

\vspace{3mm}

\unitlength1.mm
\begin{picture}(150,8.5)
\put(0,3){
\unitlength1.mm
\begin{picture}(8.25,2)
\linethickness{0.6mm}
\put(0.75,1){\line(1,0){5.5}}
\put(7.25,1){\circle{2}}
\put(0.75,1){\circle*{1.5}}
\end{picture}  }
\put(33.5,3){
\unitlength1.mm
\begin{picture}(8.25,2)
\linethickness{0.15mm}
\put(6,1){\vector(-1,0){3.75}}
\put(0.5,1){\line(1,0){1.75}}
\put(7,1){\circle{2}}
\put(0.5,1){\circle*{1.5}}
\end{picture}  }
\put(51.5,3){
\unitlength1.mm
\begin{picture}(8.25,2)
\linethickness{0.15mm}
\put(0.5,1){\vector(1,0){3.75}}
\put(6,1){\line(-1,0){1.75}}
\put(7,1){\circle{2}}
\put(0.5,1){\circle*{1.5}}
\end{picture}   }
\put(0,12.5){
\parbox[t]{15cm}{
\begin{eqnarray}
\label{S8}
=\hspace{2cm}\times\hspace{1.5cm}&=&\psi_{1,2}(x+e_1)\ \psi_{2,2}(x+e_1)
\hspace{1.4cm}
\end{eqnarray}  }}
\end{picture} 

\vspace{5mm}

Using above pictorial language for the remaining fields and 
their combinations on the odd sublattice we can gain now
already some intuitive understanding for the structures 
appearing on the r.h.s.\ of eqs.\ (\ref{V2})-(\ref{V20}). 
We give some characteristic examples, similar relations can
also be obtained for the vertices omitted now. Please note,
that the symbols on the l.h.s.\ below stand for the original 
expressions (\ref{V2}), (\ref{V5}), (\ref{V9}), (\ref{V10}), (\ref{V8})
restricted to the $\psi$, $\chi$ factors (The numerical factors
have been included with their roots, therefore.).\\

Vertex 3
\nopagebreak

\vspace{5mm}
\nopagebreak

\unitlength1.mm
\begin{picture}(150,15)
\put(0,0){
\unitlength1.mm
\begin{picture}(15,15)
\linethickness{0.15mm}
\put(7.5,12){\line(0,1){1}}
\put(7.5,10){\line(0,1){1}}
\put(7.5, 8){\line(0,1){1}}
\put(7.5, 6){\line(0,1){1}}
\put(7.5, 4){\line(0,1){1}}
\put(7.5, 2){\line(0,1){1}}
\linethickness{0.6mm}
\put(2,7.5){\line(1,0){11}}
\put(7.5,7.5){\circle*{1.5}}
\end{picture} }
\put(38,6.5){
\unitlength1.mm
\begin{picture}(15,2)
\linethickness{0.15mm}
\put(2,1){\vector(1,0){3.75}}
\put(7.5,1){\line(-1,0){1.75}}
\put(7.5,1){\vector(1,0){3.75}}
\put(13,1){\line(-1,0){1.75}}
\put(1,1){\circle{2}}
\put(7.5,1){\circle*{1.5}}
\put(14,1){\circle{2}}
\end{picture}  }
\put(63,6.5){
\unitlength1.mm
\begin{picture}(15,2)
\linethickness{0.15mm}
\put(13,1){\vector(-1,0){3.75}}
\put(7.5,1){\line(1,0){1.75}}
\put(7.5,1){\vector(-1,0){3.75}}
\put(2,1){\line(1,0){1.75}}
\put(14,1){\circle{2}}
\put(7.5,1){\circle*{1.5}}
\put(1,1){\circle{2}}
\end{picture}   }
\put(101.5,6.5){
\unitlength1.mm
\begin{picture}(8.25,2)
\linethickness{0.6mm}
\put(2,1){\line(1,0){5.5}}
\put(1,1){\circle{2}}
\put(7.5,1){\circle*{1.5}}
\end{picture}  }
\put(126.75,6.5){
\unitlength1.mm
\begin{picture}(8.25,2)
\linethickness{0.6mm}
\put(0.75,1){\line(1,0){5.5}}
\put(7.25,1){\circle{2}}
\put(0.75,1){\circle*{1.5}}
\end{picture}  }
\put(0,16){
\parbox[t]{15cm}{
\begin{eqnarray}
\label{R1}
&&=\hspace{3.1cm}\times\hspace{2.5cm}=\ \ -
\hspace{2.05cm}\times\hspace{1.3cm}
\end{eqnarray}  }}
\end{picture} 

\vspace{3mm}

Vertex 6
\nopagebreak

\vspace{5mm}
\nopagebreak

\unitlength1.mm
\begin{picture}(150,15)
\put(0,0){
\unitlength1.mm
\begin{picture}(15,15)
\linethickness{0.15mm}
\put( 6,7.5){\line(1,0){1}}
\put( 4,7.5){\line(1,0){1}}
\put( 2,7.5){\line(1,0){1}}
\put(7.5, 6){\line(0,1){1}}
\put(7.5, 4){\line(0,1){1}}
\put(7.5, 2){\line(0,1){1}}
\linethickness{0.6mm}
\put(7.5,7.5){\line(1,0){5.5}}
\put(7.5,7.5){\line(0,1){5.5}}
\put(7.5,7.5){\circle*{1.5}}
\end{picture} }
\put(44.5,6.5){
\unitlength1.mm
\begin{picture}(8.5,8.5)
\linethickness{0.15mm}
\put(1,6.5){\vector(0,-1){3.75}}
\put(1,1){\line(0,1){1.75}}
\put(1,1){\vector(1,0){3.75}}
\put(6.5,1){\line(-1,0){1.75}}
\put(1,7.5){\circle{2}}
\put(1,1){\circle*{1.5}}
\put(7.5,1){\circle{2}}
\end{picture}  }
\put(69.5,6.5){
\unitlength1.mm
\begin{picture}(8.5,8.5)
\linethickness{0.15mm}
\put(6.5,1){\vector(-1,0){3.75}}
\put(1,1){\line(1,0){1.75}}
\put(1,1){\vector(0,1){3.75}}
\put(1,6.5){\line(0,-1){1.75}}
\put(7.5,1){\circle{2}}
\put(1,1){\circle*{1.5}}
\put(1,7.5){\circle{2}}
\end{picture}  }
\put(108,6.5){
\unitlength1.mm
\begin{picture}(2,8.25)
\linethickness{0.6mm}
\put(1,0.75){\line(0,1){5.5}}
\put(1,7.25){\circle{2}}
\put(1,0.75){\circle*{1.5}}
\end{picture}  }
\put(126.75,6.5){
\unitlength1.mm
\begin{picture}(8.25,2)
\linethickness{0.6mm}
\put(0.5,1){\line(1,0){5.5}}
\put(7,1){\circle{2}}
\put(0.5,1){\circle*{1.5}}
\end{picture}  }
\put(0,16.5){
\parbox[t]{15cm}{
\begin{eqnarray}
\label{R2}
&&=\ \ {1\over 2}
\hspace{2.55cm}\times\hspace{2.5cm}=\ -{1\over 2}
\hspace{1.9cm}\times\hspace{1.3cm}
\end{eqnarray}  }}
\end{picture}

\vspace{3mm} 

Vertex 10
\nopagebreak

\vspace{5mm}
\nopagebreak

\unitlength1.mm
\begin{picture}(150,15)
\put(0,0){
\unitlength1.mm
\begin{picture}(15,15)
\linethickness{0.15mm}
\put(2,7.5){\line(1,0){11}}
\put(7.5, 6){\line(0,1){1}}
\put(7.5, 4){\line(0,1){1}}
\put(7.5, 2){\line(0,1){1}}
\linethickness{0.6mm}
\put(7.5,7.5){\line(0,1){5.5}}
\put(7.5,7.5){\circle*{1.5}}
\end{picture}  }
\put(38,6.5){
\unitlength1.mm
\begin{picture}(8.5,8.5)
\linethickness{0.15mm}
\put(7.5,1){\line(-1,0){5.5}}
\put(7.5,1){\line(0,1){5.5}}
\put(7.5,7.5){\circle{2}}
\put(7.5,1){\circle*{1.5}}
\put(1,1){\circle{2}}
\end{picture} }
\put(69.5,6.5){
\unitlength1.mm
\begin{picture}(8.5,8.5)
\linethickness{0.15mm}
\put(1,1){\line(0,1){5.5}}
\put(1,1){\line(1,0){5.5}}
\put(1,7.5){\circle{2}}
\put(1,1){\circle*{1.5}}
\put(7.5,1){\circle{2}}
\end{picture} }
\put(101.5,6.5){
\unitlength1.mm
\begin{picture}(15,2)
\linethickness{0.15mm}
\put(2,1){\line(1,0){11}}
\put(1,1){\circle{2}}
\put(7.5,1){\circle*{1.5}}
\put(14,1){\circle{2}}
\end{picture}  }
\put(126.75,6.75){
\unitlength1.mm
\begin{picture}(2,8.25)
\linethickness{0.6mm}
\put(1,0.75){\line(0,1){5.5}}
\put(1,7.25){\circle{2}}
\put(1,0.75){\circle*{1.5}}
\end{picture}  }
\put(0,16.5){
\parbox[t]{15cm}{
\begin{eqnarray}
\label{R4}
&&=\ -{1\over 2}
\hspace{2.35cm}\times\hspace{2.5cm}=\ \ {1\over 2}
\hspace{2.05cm}\times\hspace{1.3cm}
\end{eqnarray}  }}
\end{picture}

\vspace{3mm} 

Vertex 14
\nopagebreak

\vspace{5mm}
\nopagebreak

\unitlength1.mm
\begin{picture}(150,15)
\put(0,0){
\unitlength1.mm
\begin{picture}(15,15)
\linethickness{0.15mm}
\put(7.5, 6){\line(0,1){1}}
\put(7.5, 4){\line(0,1){1}}
\put(7.5, 2){\line(0,1){1}}
\put(2,7.5){\line(1,0){5.5}}
\put(7.5,7.5){\line(0,1){5.5}}
\linethickness{0.6mm}
\put(7.5,7.5){\line(1,0){5.5}}
\put(7.5,7.5){\circle*{1.5}}
\end{picture}  }
\put(38,6.5){
\unitlength1.mm
\begin{picture}(15,2)
\linethickness{0.15mm}
\put(2,1){\line(1,0){11}}
\put(1,1){\circle{2}}
\put(7.5,1){\circle*{1.5}}
\put(14,1){\circle{2}}
\end{picture}  }
\put(69.5,6.5){
\unitlength1.mm
\begin{picture}(8.5,8.5)
\linethickness{0.15mm}
\put(1,1){\line(0,1){5.5}}
\put(1,1){\line(1,0){5.5}}
\put(1,7.5){\circle{2}}
\put(1,1){\circle*{1.5}}
\put(7.5,1){\circle{2}}
\end{picture} }
\put(101.5,6.5){
\unitlength1.mm
\begin{picture}(8.5,8.5)
\linethickness{0.15mm}
\put(7.5,1){\line(-1,0){5.5}}
\put(7.5,1){\line(0,1){5.5}}
\put(7.5,7.5){\circle{2}}
\put(7.5,1){\circle*{1.5}}
\put(1,1){\circle{2}}
\end{picture} }
\put(126.75,6.5){
\unitlength1.mm
\begin{picture}(8.25,2)
\linethickness{0.6mm}
\put(0.75,1){\line(1,0){5.5}}
\put(7.25,1){\circle{2}}
\put(0.75,1){\circle*{1.5}}
\end{picture}  }
\put(0,16.5){
\parbox[t]{15cm}{
\begin{eqnarray}
\label{R5}
&&=\ {-1\over\sqrt{2}}
\hspace{2.25cm}\times\hspace{2.5cm}=\ \ {1\over\sqrt{2}}
\hspace{1.65cm}\times\hspace{1.3cm}
\end{eqnarray}  }}
\end{picture}

\vspace{3mm} 

Vertex 20
\nopagebreak

\vspace{5mm}
\nopagebreak

\unitlength1.mm
\begin{picture}(150,15)
\put(0,0){
\unitlength1.mm
\begin{picture}(15,15)
\linethickness{0.15mm}
\put(2,7.5){\line(1,0){11}}
\put(7.5,2){\line(0,1){11}}
\put(7.5,7.5){\circle*{1.5}}
\end{picture} }
\put(38,6.5){
\unitlength1.mm
\begin{picture}(8.5,8.5)
\linethickness{0.15mm}
\put(7.5,1){\line(-1,0){5.5}}
\put(7.5,1){\line(0,1){5.5}}
\put(7.5,7.5){\circle{2}}
\put(7.5,1){\circle*{1.5}}
\put(1,1){\circle{2}}
\end{picture} }
\put(69.5,0){
\unitlength1.mm
\begin{picture}(8.5,8.5)
\linethickness{0.15mm}
\put(1,7.5){\line(1,0){5.5}}
\put(1,7.5){\line(0,-1){5.5}}
\put(7.5,7.5){\circle{2}}
\put(1,7.5){\circle*{1.5}}
\put(1,1){\circle{2}}
\end{picture}  }
\put(101.5,0){
\unitlength1.mm
\begin{picture}(8.5,8.5)
\linethickness{0.15mm}
\put(7.5,7.5){\line(0,-1){5.5}}
\put(7.5,7.5){\line(-1,0){5.5}}
\put(1,7.5){\circle{2}}
\put(7.5,7.5){\circle*{1.5}}
\put(7.5,1){\circle{2}}
\end{picture}  }
\put(126.75,6.5){
\unitlength1.mm
\begin{picture}(8.5,8.5)
\linethickness{0.15mm}
\put(1,1){\line(0,1){5.5}}
\put(1,1){\line(1,0){5.5}}
\put(1,7.5){\circle{2}}
\put(1,1){\circle*{1.5}}
\put(7.5,1){\circle{2}}
\end{picture} }
\put(0,16.5){
\parbox[t]{15cm}{
\begin{eqnarray}
\label{R3}
&&=\ \ {1\over 2}
\hspace{2.55cm}\times\hspace{2.5cm}-\ \ {1\over 2}
\hspace{2.05cm}\times\hspace{1.3cm}
\end{eqnarray}  }}
\end{picture}

\vspace{5mm}

One immediately recognizes the significance of the numerical 
factors for each vertex. The rule simply is that each corner
element (\ref{S5})-(\ref{S6}) is associated with a factor $1/\sqrt{2}$.  
Although more useful at the end, in this respect the 
graphical representations given rightmost in eqs.\ (\ref{R2}), 
(\ref{R4}) are somewhat misleading.\\

\parindent1.5em

What remains to be done now is to discuss the result for the 
Grassmann integration on the odd sublattice. For the moment we 
exclude vertex 20 (and its equivalent on the odd sublattice)
 from the consideration as its complexity requires special 
attention. It will be discussed at a later stage. 
In view of the vertex arrow rule one finds the following non-zero 
results (We give only few typical cases,
all others are related by symmetry considerations of the 
pictures using rotations and reflections.). The graphical symbols
below denote in an intuitive way the product of (four) Grassmann 
variables, their ordering performed clockwise with respect to 
the points on the even sublattice starting with the 
leftmost point in a picture. The black dot in 
the hollow circle of the point on the odd sublattice indicates
that the Grassmann integration has been performed now.\\

\parindent0.em

\unitlength1.mm
\begin{picture}(150,15)
\put(14,0){
\unitlength1.mm
\begin{picture}(14.5,2)
\linethickness{0.6mm}
\put(0.75,1){\line(1,0){5.5}}
\put(8.25,1){\line(1,0){5.5}}
\put(7.25,1){\circle{2}}
\put(7.25,1){\circle*{1.5}}
\put(0.75,1){\circle*{1.5}}
\put(13.75,1){\circle*{1.5}}
\end{picture}  }
\put(0,11.5){
\parbox[t]{15cm}{
\begin{eqnarray}
\label{DI1}
&&\hspace{1.5cm}=\ \int\prod_{i=1}^2\prod_{\alpha=1}^2
d\psi_{i,\alpha}(x)\ 
\psi_{1,2}(x)\ \psi_{2,2}(x)\ \psi_{1,1}(x)\ \psi_{2,1}(x)\ =
\ \ -1
\end{eqnarray}  }}
\end{picture}

\unitlength1.mm
\begin{picture}(150,15)
\put(20.25,0){
\unitlength1.mm
\begin{picture}(8.25,8.25)
\linethickness{0.6mm}
\put(2,1){\line(1,0){5.5}}
\put(1,2){\line(0,1){5.5}}
\put(1,1){\circle{2}}
\put(1,1){\circle*{1.5}}
\put(7.5,1){\circle*{1.5}}
\put(1,7.5){\circle*{1.5}}
\end{picture}  }
\put(0,10.5){
\parbox[t]{15cm}{
\begin{eqnarray}
\label{DI2}
&&\hspace{3.8cm}=\ -{1\over 2}\ \ \hspace{3.2cm}
\end{eqnarray}  }}
\end{picture} 

\unitlength1.mm
\begin{picture}(150,15)
\put(14,0){
\unitlength1.mm
\begin{picture}(14.5,8.25)
\linethickness{0.6mm}
\put(7.25,2){\line(0,1){5.5}}
\put(7.25,1){\circle{2}}
\put(7.25,1){\circle*{1.5}}
\put(7.25,7.5){\circle*{1.5}}
\put(0.75,1){\circle*{1.5}}
\put(13.75,1){\circle*{1.5}}
\linethickness{0.15mm}
\put(13.75,1){\vector(-1,0){3.75}}
\put(8.25,1){\line(1,0){1.75}}
\put(6.25,1){\vector(-1,0){3.75}}
\put(0.75,1){\line(1,0){1.75}}
\end{picture}  }
\put(50,0){
\unitlength1.mm
\begin{picture}(14.5,8.25)
\linethickness{0.6mm}
\put(7.25,2){\line(0,1){5.5}}
\put(7.25,1){\circle{2}}
\put(7.25,1){\circle*{1.5}}
\put(7.25,7.5){\circle*{1.5}}
\put(0.75,1){\circle*{1.5}}
\put(13.75,1){\circle*{1.5}}
\linethickness{0.15mm}
\put(8.25,1){\vector(1,0){3.75}}
\put(13.75,1){\line(-1,0){1.75}}
\put(0.75,1){\vector(1,0){3.75}}
\put(6.25,1){\line(-1,0){1.75}}
\end{picture}  }
\put(0,10.5){
\parbox[t]{15cm}{
\begin{eqnarray}
\label{DI3}
&&=\ -\hspace{3.0cm}=\ \ {1\over 2}\ \ \hspace{3.5cm}
\end{eqnarray}  }}
\end{picture} 

\unitlength1.mm
\begin{picture}(150,15)
\put(14,0){
\unitlength1.mm
\begin{picture}(14.5,8.25)
\linethickness{0.6mm}
\put(8.25,1){\line(1,0){5.5}}
\put(7.25,1){\circle{2}}
\put(7.25,1){\circle*{1.5}}
\put(0.75,1){\circle*{1.5}}
\put(13.75,1){\circle*{1.5}}
\put(7.25,7.5){\circle*{1.5}}
\linethickness{0.15mm}
\put(7.25,2){\vector(0,1){3.75}}
\put(7.25,7.5){\line(0,-1){1.75}}
\put(0.75,1){\vector(1,0){3.75}}
\put(6.25,1){\line(-1,0){1.75}}
\end{picture}  }
\put(50,0){
\unitlength1.mm
\begin{picture}(14.5,8.25)
\linethickness{0.6mm}
\put(8.25,1){\line(1,0){5.5}}
\put(7.25,1){\circle{2}}
\put(7.25,1){\circle*{1.5}}
\put(0.75,1){\circle*{1.5}}
\put(13.75,1){\circle*{1.5}}
\put(7.25,7.5){\circle*{1.5}}
\linethickness{0.15mm}
\put(7.25,7.5){\vector(0,-1){3.75}}
\put(7.25,2){\line(0,1){1.75}}
\put(6.25,1){\vector(-1,0){3.75}}
\put(0.75,1){\line(1,0){1.75}}
\end{picture}  }
\put(0,10.5){
\parbox[t]{15cm}{
\begin{eqnarray}
\label{DI4}
&&=\ -\hspace{3.0cm}=\ {1\over\sqrt{2}}\hspace{3.5cm}
\end{eqnarray}  }}
\end{picture} 

\vspace{5mm}

 From these results one recognizes that links with $k_l = 1$ (thin lines)
must form self-avoiding loops. It remains to determine the weight of such 
loops inasmuch as vertices contributing to such loops have two terms each
(see, e.g., eqs.\ (\ref{R4}), (\ref{R5}) and (\ref{S1}), (\ref{S4})). 
However, due to the vertex arrow rule 
only those products of terms give a non-vanishing contribution which 
form an oriented loop. This leaves for each loop just two terms
(corresponding to the two possible orientations of the loop). 
Both of these terms give exactly the same contribution. This can 
be seen easily by interchanging flavours 1 and 2 what amounts to
arrow reversal. Each of the 
building blocks of a thin line ($k_l = 1$) 
loop (\ref{S1})-(\ref{S6}) transforms
under flavour exchange into the negative of itself. Also, in view of 
eqs.\ (\ref{DI3}), (\ref{DI4}) at each point belonging to the odd 
sublattice of a given thin line 
loop arrow reversal brings a relative 
minus sign about. Now, these two minus signs match locally at each point
of the odd sublattice of a given thin line
loop. Consequently, the weight
of each thin line loop is given by the product of the weights of 
its constituting vertices multiplied by a factor of 2 (the loop 
multiplicity). Finally, one 
recognizes from eqs.\ (\ref{DI2})-(\ref{DI4}) that each corner (being
the product of 2 fields originally) contributes a factor of $1/\sqrt{2}$
(For eq.\ (\ref{DI3}) one has to apply an interpretation analogous
to the middle part of eq.\ (\ref{R4}), of course.).\\

\parindent1.5em
Let us now extend the analysis by also allowing configurations
containing vertex 20 we omitted so far. 
It is clear from the beginning that the picture
of self-avoiding thin line ($k_l = 1$) loops cannot be maintained in any 
naive way, but we will see how it can still be used in 
a modified version. The above expression (\ref{R3}) already suggests
what can be done. One may identify the vertex 20 with two loop segments
passing each other in a given point without intersection. Clearly, there
are two ways of doing so which are reflected in the two different terms
shown in eq.\ (\ref{R3}). Each of these two terms corresponds to 
four possible combinations of segments of oriented loops (each being 
related to the product of four fields on the odd sublattice, 
cf.\ eqs.\ (\ref{S5})-(\ref{S6})). 
If the two segments happen to be just different 
parts of one loop two of the four possible combinations 
of oriented thin line loop segments are
effective in the final Grassmann integration over the odd sublattice only
(The other two combinations must then lead to a zero result according to the 
vertex arrow rule.). If they belong to two different loops all
terms contribute at the end. In considering the two terms in eq.\ 
(\ref{R3}), it is clear however that, if one term is part of two
different loops, the other must just belong to one loop returning 
at some stage to the point $x\in\Lambda_e$ under consideration. 
Consequently, the two terms in eq.\ (\ref{R3}) differ in their
contribution at the end just by a factor of 2 if it comes to the 
loop count (just for geometrical reasons). 
As each self-avoiding loop contributes a factor of 2
(related to the two different orientations of the loop) the 
term which is part of two different loops will be related to a factor of 4
while the other one is related to a factor of 2 only. The remaining
worry is what can be said with respect to the relative sign of 
the contributions from the two terms in eq.\ (\ref{R3}). We will see
that the relative minus sign in eq.\ (\ref{R3}) is spurious (arising
 from ordering effects of Grassmann variables) and that in fact the 
contributions from the two terms in eq.\ (\ref{R3}) add up at the 
end. This can be detected by noting that if one writes the two 
geometrically different terms in eq.\ (\ref{R3}) as sum over 
products of (four) fields two of these products are the same, namely
$\psi_{1,1}(x-e_1)\ \chi_{2,2}(x+e_2)\ \psi_{1,2}(x+e_1)\ \chi_{2,1}(x-e_2)$ 
(The horizontal arrows are flowing in into the point $x$
while the vertical ones are flowing out.) and 
$\psi_{2,1}(x-e_1)\ \chi_{1,2}(x+e_2)\ \psi_{2,2}(x+e_1)\ \chi_{1,1}(x-e_2)$
(The horizontal arrows are flowing out of the point $x$
while the vertical ones are flowing in.).
Both contribute to the l.h.s.\ of eq.\ (\ref{R3}) with a total factor 
of $1$ at the end. This proves that contributions from both 
terms in eq.\ (\ref{R3}) add up after properly taking into account ordering
effects.\\

Finally, to complete the picture we have to discuss 
the analogue of vertex 20 on the 
odd sublattice. One finds the following non-vanishing 
Grassmann integrals beyond those given in eqs.\ (\ref{DI1})-(\ref{DI4})
(We apply here the same conventions as there.).\\

\parindent0.em

\unitlength1.mm
\begin{picture}(150,14.5)
\put(3,0){
\unitlength1.mm
\begin{picture}(14.5,14.5)
\linethickness{0.15mm}
\put(0.75,7.25){\vector(1,0){3.75}}
\put(6.25,7.25){\line(-1,0){1.75}}
\put(13.75,7.25){\vector(-1,0){3.75}}
\put(8.25,7.25){\line(1,0){1.75}}
\put(7.25,6.25){\vector(0,-1){3.75}}
\put(7.25,0.75){\line(0,1){1.75}}
\put(7.25,8.25){\vector(0,1){3.75}}
\put(7.25,13.75){\line(0,-1){1.75}}
\put(7.25,7.25){\circle{2}}
\put(7.25,7.25){\circle*{1.5}}
\put(13.75,7.25){\circle*{1.5}}
\put(0.75,7.25){\circle*{1.5}}
\put(7.25,13.75){\circle*{1.5}}
\put(7.25,0.75){\circle*{1.5}}
\end{picture} }
\put(33,0){
\unitlength1.mm
\begin{picture}(14.5,14.5)
\linethickness{0.15mm}
\put(8.25,7.25){\vector(1,0){3.75}}
\put(13.75,7.25){\line(-1,0){1.75}}
\put(6.25,7.25){\vector(-1,0){3.75}}
\put(0.75,7.25){\line(1,0){1.75}}
\put(7.25,13.75){\vector(0,-1){3.75}}
\put(7.25,8.25){\line(0,1){1.75}}
\put(7.25,0.75){\vector(0,1){3.75}}
\put(7.25,6.25){\line(0,-1){1.75}}
\put(7.25,7.25){\circle{2}}
\put(7.25,7.25){\circle*{1.5}}
\put(13.75,7.25){\circle*{1.5}}
\put(0.75,7.25){\circle*{1.5}}
\put(7.25,13.75){\circle*{1.5}}
\put(7.25,0.75){\circle*{1.5}}
\end{picture}  }
\put(0,15.75){
\parbox[t]{15cm}{
\begin{eqnarray}
\label{ver8-1}
&&=\hspace{9.cm}=\ -1\hspace{6mm}
\end{eqnarray}  }}
\end{picture} 

\vspace{5mm}

\unitlength1.mm
\begin{picture}(150,14.5)
\put(3,0){
\unitlength1.mm
\begin{picture}(14.5,14.5)
\linethickness{0.15mm}
\put(8.25,7.25){\vector(1,0){3.75}}
\put(13.75,7.25){\line(-1,0){1.75}}
\put(0.75,7.25){\vector(1,0){3.75}}
\put(6.25,7.25){\line(-1,0){1.75}}
\put(7.25,8.25){\vector(0,1){3.75}}
\put(7.25,13.75){\line(0,-1){1.75}}
\put(7.25,0.75){\vector(0,1){3.75}}
\put(7.25,6.25){\line(0,-1){1.75}}
\put(7.25,7.25){\circle{2}}
\put(7.25,7.25){\circle*{1.5}}
\put(13.75,7.25){\circle*{1.5}}
\put(0.75,7.25){\circle*{1.5}}
\put(7.25,13.75){\circle*{1.5}}
\put(7.25,0.75){\circle*{1.5}}
\end{picture} }
\put(33,0){
\unitlength1.mm
\begin{picture}(14.5,14.5)
\linethickness{0.15mm}
\put(13.75,7.25){\vector(-1,0){3.75}}
\put(8.25,7.25){\line(1,0){1.75}}
\put(6.25,7.25){\vector(-1,0){3.75}}
\put(0.75,7.25){\line(1,0){1.75}}
\put(7.25,13.75){\vector(0,-1){3.75}}
\put(7.25,8.25){\line(0,1){1.75}}
\put(7.25,6.25){\vector(0,-1){3.75}}
\put(7.25,0.75){\line(0,1){1.75}}
\put(7.25,7.25){\circle{2}}
\put(7.25,7.25){\circle*{1.5}}
\put(13.75,7.25){\circle*{1.5}}
\put(0.75,7.25){\circle*{1.5}}
\put(7.25,13.75){\circle*{1.5}}
\put(7.25,0.75){\circle*{1.5}}
\end{picture} }
\put(63,0){
\unitlength1.mm
\begin{picture}(14.5,14.5)
\linethickness{0.15mm}
\put(13.75,7.25){\vector(-1,0){3.75}}
\put(8.25,7.25){\line(1,0){1.75}}
\put(6.25,7.25){\vector(-1,0){3.75}}
\put(0.75,7.25){\line(1,0){1.75}}
\put(7.25,8.25){\vector(0,1){3.75}}
\put(7.25,13.75){\line(0,-1){1.75}}
\put(7.25,0.75){\vector(0,1){3.75}}
\put(7.25,6.25){\line(0,-1){1.75}}
\put(7.25,7.25){\circle{2}}
\put(7.25,7.25){\circle*{1.5}}
\put(13.75,7.25){\circle*{1.5}}
\put(0.75,7.25){\circle*{1.5}}
\put(7.25,13.75){\circle*{1.5}}
\put(7.25,0.75){\circle*{1.5}}
\end{picture}  }
\put(93,0){
\unitlength1.mm
\begin{picture}(14.5,14.5)
\linethickness{0.15mm}
\put(8.25,7.25){\vector(1,0){3.75}}
\put(13.75,7.25){\line(-1,0){1.75}}
\put(0.75,7.25){\vector(1,0){3.75}}
\put(6.25,7.25){\line(-1,0){1.75}}
\put(7.25,13.75){\vector(0,-1){3.75}}
\put(7.25,8.25){\line(0,1){1.75}}
\put(7.25,6.25){\vector(0,-1){3.75}}
\put(7.25,0.75){\line(0,1){1.75}}
\put(7.25,7.25){\circle{2}}
\put(7.25,7.25){\circle*{1.5}}
\put(13.75,7.25){\circle*{1.5}}
\put(0.75,7.25){\circle*{1.5}}
\put(7.25,13.75){\circle*{1.5}}
\put(7.25,0.75){\circle*{1.5}}
\end{picture} }
\put(0,16.25){
\parbox[t]{15cm}{
\begin{eqnarray}
\label{ver8-2}
&&=\hspace{2.7cm}=\hspace{2.7cm}=\hspace{3cm}=\ \ {1\over 2}\hspace{6mm}
\end{eqnarray}  }}
\end{picture} 

\vspace{5mm}

Again, the loop picture is helpful to recognize the underlying 
structure. Each of the vertices given in eq.\ (\ref{ver8-2})
allows in an unique way an interpretation as standing for 
two segments of oriented loops passing each other in the given
point on the odd sublattice without intersection (The first two 
diagrams are standing in correspondence to the first term in eq.\ (\ref{R3}) 
while the other two are being analogous to the second term in 
eq.\ (\ref{R3}).). This interpretation is supported by the 
appearance of the factor $1/2$ (characteristic for two corners)  
on the r.h.s.\ of eq.\ (\ref{ver8-2}). Arrow reversal for
just one loop corner transforms any diagram of eq.\  (\ref{ver8-2})
into one of the diagrams shown in eq.\  (\ref{ver8-1}). This 
immediately sheds light onto the different numerical result for these.
We may imagine each diagram in eq.\ (\ref{ver8-1}) being a sum
of two identical terms each of which is supplied with 
an additional specification. The additional specification is
provided by one of the two possible ways of two oriented loop 
segments passing each other without intersection in the given lattice site
(The point is that the diagrams in eq.\ (\ref{ver8-1}) in 
opposition to those in eq.\ (\ref{ver8-2}) allow two 
different interpretations each.). Each of those terms with the additional 
specification contributes $\ -1/2$ then. In analogy with eq.\ 
(\ref{DI4}), under arrow reversal for just one oriented loop corner  
we have obtained this way a partner for each diagram in 
eq.\ (\ref{ver8-2}). This partner diagram has as weight 
the negative of the corresponding diagram in eq.\ (\ref{ver8-2}). 
This completes the discussion.\\

\parindent1.5em

The above consideration allows us to use the following 
efficient graphical rule
to determine the weight of any cluster containing vertex 20 (and 
its analogue on the odd sublattice).\\

\parindent0.em

\unitlength1.mm
\begin{picture}(150,15)
\put(10,0){
\unitlength1.mm
\begin{picture}(15,15)
\linethickness{0.15mm}
\put(2,7.5){\line(1,0){11}}
\put(7.5,2){\line(0,1){11}}
\end{picture} }
\put(50,0){
\unitlength1.mm
\begin{picture}(15,15)
\linethickness{0.15mm}
\put(2,7.5){\line(1,0){3.5}}
\put(9.5,7.5){\line(1,0){3.5}}
\put(5.5,9.5){\oval(4,4)[br]}
\put(9.5,5.5){\oval(4,4)[tl]}
\put(7.5,2){\line(0,1){3.5}}
\put(7.5,9.5){\line(0,1){3.5}}
\end{picture} }
\put(80,0){
\unitlength1.mm
\begin{picture}(15,15)
\linethickness{0.15mm}
\put(2,7.5){\line(1,0){3.5}}
\put(9.5,7.5){\line(1,0){3.5}}
\put(9.5,9.5){\oval(4,4)[bl]}
\put(5.5,5.5){\oval(4,4)[tr]}
\put(7.5,2){\line(0,1){3.5}}
\put(7.5,9.5){\line(0,1){3.5}}
\end{picture} }
\put(0,16){
\parbox[t]{15cm}{
\begin{eqnarray}
\label{rule}
&&\longrightarrow\hspace{3cm}+\hspace{4.8cm}
\end{eqnarray}  }}
\end{picture}

\vspace{5mm}

In words, the rule shown above simply requires to replace any vertex 20
(and its analogue on the odd sublattice) within a given
cluster by two vertices built of two thin line loop segments
passing each other without intersection. This leads to 
$2^n$ different cluster configurations when $n$ is the number
of times vertex 20 (and its analogue on the odd sublattice)
occurs in the original configuration.
The investigation of the $\psi$, $\chi$ subsystem shows
that this rule can reproduce the contribution of vertex 20
(and its analogue on the odd sublattice) if one assigns each of
the two new vertices a factor $1/2$, counts any  
thin line ($k_l = 1$) loop in the newly created configurations
with a factor (loop multiplicity) of 2 (characteristic for oriented loops),
and then adds up the weights of all these configurations. 
As the loop count gives different results 
in a given geometrical situation for the two vertices appearing on 
the r.h.s.\ of eq.\ (\ref{rule}) the generalization of the above rule
to the whole $\bar\psi$, $\bar\chi$, $\psi$, $\chi$ system
requires further thought. As the rule (\ref{rule}) can be 
applied for the $\bar\psi$, $\bar\chi$ and the  $\psi$, $\chi$ 
subsystems separately each occurrence of a vertex 20 (and its 
analogue on the odd sublattice) leads to a situation where 
it is met twice in a product of field combinations related 
to this lattice site as one recognizes from the picture below
(The vertices of the $\bar\psi$, $\bar\chi$ subsystem (upper line)
are emphasized by stronger lines. If one wants to have in mind
some explicit expression for the intuitive picture below, 
eq.\ (\ref{R3}) is recommended as reference.).\\

\unitlength1.mm
\begin{picture}(150,35)
\put(50,0){
\unitlength1.mm
\begin{picture}(15,15)
\linethickness{0.15mm}
\put(2,7.5){\line(1,0){3.5}}
\put(9.5,7.5){\line(1,0){3.5}}
\put(5.5,9.5){\oval(4,4)[br]}
\put(9.5,5.5){\oval(4,4)[tl]}
\put(7.5,2){\line(0,1){3.5}}
\put(7.5,9.5){\line(0,1){3.5}}
\end{picture} }
\put(80,0){
\unitlength1.mm
\begin{picture}(15,15)
\linethickness{0.15mm}
\put(2,7.5){\line(1,0){3.5}}
\put(9.5,7.5){\line(1,0){3.5}}
\put(9.5,9.5){\oval(4,4)[bl]}
\put(5.5,5.5){\oval(4,4)[tr]}
\put(7.5,2){\line(0,1){3.5}}
\put(7.5,9.5){\line(0,1){3.5}}
\end{picture} }
\put(50,20){
\unitlength1.mm
\begin{picture}(15,15)
\linethickness{0.3mm}
\put(2,7.5){\line(1,0){3.5}}
\put(9.5,7.5){\line(1,0){3.5}}
\put(5.5,9.5){\oval(4,4)[br]}
\put(9.5,5.5){\oval(4,4)[tl]}
\put(7.5,2){\line(0,1){3.5}}
\put(7.5,9.5){\line(0,1){3.5}}
\end{picture} }
\put(80,20){
\unitlength1.mm
\begin{picture}(15,15)
\linethickness{0.3mm}
\put(2,7.5){\line(1,0){3.5}}
\put(9.5,7.5){\line(1,0){3.5}}
\put(9.5,9.5){\oval(4,4)[bl]}
\put(5.5,5.5){\oval(4,4)[tr]}
\put(7.5,2){\line(0,1){3.5}}
\put(7.5,9.5){\line(0,1){3.5}}
\end{picture} }
\put(0,37){
\parbox[t]{15cm}{
\begin{eqnarray}
\label{rulesquared}
&&\hspace{7.5mm}\Bigg(\hspace{3cm}+\hspace{3cm}\Bigg)\hspace{1.8cm}
\nonumber\\[0.8cm]
&&\times\ \ \ \Bigg(\hspace{3cm}+\hspace{3cm}\Bigg)\hspace{1.8cm}
\end{eqnarray}  }}
\end{picture}

\vspace{5mm}

There does not exist any problem connected with the two terms where the 
vertices are identical in the $\bar\psi$, $\bar\chi$ and 
the  $\psi$, $\chi$ subsystems. However, there are 
also two cross terms which have to be taken care of.
In principle, it would be our aim to apply the rule (\ref{rule}) to 
the complete $\bar\psi$, $\bar\chi$, $\psi$, $\chi$ system
by simply changing the thin line loop multiplicity from 2 (for each 
subsystem) to 4 (for the complete system) but 
these cross terms prevent us from doing so.
Consequently, if one wants to determine the weight
of a given cluster for the full $\bar\psi$, $\bar\chi$, $\psi$, $\chi$ 
system one has to carry out the loop count for one subsystem
with loop multiplicity 2 and then to square the result to get
the correct final weight.
The study of some simple clusters suggests that one might 
be able to simplify this procedure by redefining the weights 
of the two vertices on the r.h.s.\ of eq.\ (\ref{rule})
(i.e., by hiding the contribution of the two cross terms
in eq.\ (\ref{rulesquared})
in a modified weight for the two terms where the 
vertices are identical in the $\bar\psi$, $\bar\chi$ and 
$\psi$, $\chi$ subsystems). However, closer inspection 
reveals that such a simplification is not possible.\\

\parindent1.5em

We are now in a position to characterize the vertex model 
equivalent to the two-flavour strong coupling lattice Schwinger model
with Wilson fermions. It is a (modified) 3-state 20-vertex model on the 
square lattice. The vertices are shown in Fig.\ 1 
and the vertex weights are given below.
(All other vertices of the 3-state model one might think 
of have vanishing weight.). The term 3-state
refers to the three possible states of links ($k_l=0,1,2$;
dashed, thin and thick lines respectively).

\parindent0.em

\begin{table}[t]
\unitlength1.mm
\begin{picture}(150,85)
\put(0,70){
\unitlength1.mm
\begin{picture}(15,15)
\linethickness{0.15mm}
\put(12,7.5){\line(1,0){1}}
\put(10,7.5){\line(1,0){1}}
\put( 8,7.5){\line(1,0){1}}
\put( 6,7.5){\line(1,0){1}}
\put( 4,7.5){\line(1,0){1}}
\put( 2,7.5){\line(1,0){1}}
\put(7.5,12){\line(0,1){1}}
\put(7.5,10){\line(0,1){1}}
\put(7.5, 8){\line(0,1){1}}
\put(7.5, 6){\line(0,1){1}}
\put(7.5, 4){\line(0,1){1}}
\put(7.5, 2){\line(0,1){1}}
\end{picture}  }
\put(20,70){
\unitlength1.mm
\begin{picture}(15,15)
\linethickness{0.15mm}
\put(12,7.5){\line(1,0){1}}
\put(10,7.5){\line(1,0){1}}
\put( 8,7.5){\line(1,0){1}}
\put( 6,7.5){\line(1,0){1}}
\put( 4,7.5){\line(1,0){1}}
\put( 2,7.5){\line(1,0){1}}
\linethickness{0.6mm}
\put(7.5,2){\line(0,1){11}}
\end{picture}  }
\put(40,70){
\unitlength1.mm
\begin{picture}(15,15)
\linethickness{0.15mm}
\put(7.5,12){\line(0,1){1}}
\put(7.5,10){\line(0,1){1}}
\put(7.5, 8){\line(0,1){1}}
\put(7.5, 6){\line(0,1){1}}
\put(7.5, 4){\line(0,1){1}}
\put(7.5, 2){\line(0,1){1}}
\linethickness{0.6mm}
\put(2,7.5){\line(1,0){11}}
\end{picture}  }
\put(60,70){
\unitlength1.mm
\begin{picture}(15,15)
\linethickness{0.15mm}
\put( 6,7.5){\line(1,0){1}}
\put( 4,7.5){\line(1,0){1}}
\put( 2,7.5){\line(1,0){1}}
\put(7.5,12){\line(0,1){1}}
\put(7.5,10){\line(0,1){1}}
\put(7.5, 8){\line(0,1){1}}
\linethickness{0.6mm}
\put(7.2,7.5){\line(1,0){5.8}}
\put(7.5,2){\line(0,1){5.8}}
\end{picture}  }
\put(80,70){
\unitlength1.mm
\begin{picture}(15,15)
\linethickness{0.15mm}
\put(12,7.5){\line(1,0){1}}
\put(10,7.5){\line(1,0){1}}
\put( 8,7.5){\line(1,0){1}}
\put(7.5, 6){\line(0,1){1}}
\put(7.5, 4){\line(0,1){1}}
\put(7.5, 2){\line(0,1){1}}
\linethickness{0.6mm}
\put(2,7.5){\line(1,0){5.8}}
\put(7.5,7.2){\line(0,1){5.8}}
\end{picture}  }
\put(100,70){
\unitlength1.mm
\begin{picture}(15,15)
\linethickness{0.15mm}
\put( 6,7.5){\line(1,0){1}}
\put( 4,7.5){\line(1,0){1}}
\put( 2,7.5){\line(1,0){1}}
\put(7.5, 6){\line(0,1){1}}
\put(7.5, 4){\line(0,1){1}}
\put(7.5, 2){\line(0,1){1}}
\linethickness{0.6mm}
\put(7.2,7.5){\line(1,0){5.8}}
\put(7.5,7.2){\line(0,1){5.8}}
\end{picture}  }
\put(120,70){
\unitlength1.mm
\begin{picture}(15,15)
\linethickness{0.15mm}
\put(12,7.5){\line(1,0){1}}
\put(10,7.5){\line(1,0){1}}
\put( 8,7.5){\line(1,0){1}}
\put(7.5,12){\line(0,1){1}}
\put(7.5,10){\line(0,1){1}}
\put(7.5, 8){\line(0,1){1}}
\linethickness{0.6mm}
\put(2,7.5){\line(1,0){5.8}}
\put(7.5,2){\line(0,1){5.8}}
\end{picture}  }
\put(0,40){
\unitlength1.mm
\begin{picture}(15,15)
\linethickness{0.15mm}
\put(12,7.5){\line(1,0){1}}
\put(10,7.5){\line(1,0){1}}
\put( 8,7.5){\line(1,0){1}}
\put(7.5,2){\line(0,1){11}}
\linethickness{0.6mm}
\put(2,7.5){\line(1,0){5.5}}
\end{picture}  }
\put(20,40){
\unitlength1.mm
\begin{picture}(15,15)
\linethickness{0.15mm}
\put( 6,7.5){\line(1,0){1}}
\put( 4,7.5){\line(1,0){1}}
\put( 2,7.5){\line(1,0){1}}
\put(7.5,2){\line(0,1){11}}
\linethickness{0.6mm}
\put(7.5,7.5){\line(1,0){5.5}}
\end{picture}  }
\put(40,40){
\unitlength1.mm
\begin{picture}(15,15)
\linethickness{0.15mm}
\put(2,7.5){\line(1,0){11}}
\put(7.5, 6){\line(0,1){1}}
\put(7.5, 4){\line(0,1){1}}
\put(7.5, 2){\line(0,1){1}}
\linethickness{0.6mm}
\put(7.5,7.5){\line(0,1){5.5}}
\end{picture} }
\put(60,40){
\unitlength1.mm
\begin{picture}(15,15)
\linethickness{0.15mm}
\put(2,7.5){\line(1,0){11}}
\put(7.5,12){\line(0,1){1}}
\put(7.5,10){\line(0,1){1}}
\put(7.5, 8){\line(0,1){1}}
\linethickness{0.6mm}
\put(7.5,2){\line(0,1){5.5}}
\end{picture}  }
\put(80,40){
\unitlength1.mm
\begin{picture}(15,15)
\linethickness{0.15mm}
\put(7.5,12){\line(0,1){1}}
\put(7.5,10){\line(0,1){1}}
\put(7.5, 8){\line(0,1){1}}
\put(7.5,7.5){\line(1,0){5.5}}
\put(7.5,2){\line(0,1){5.5}}
\linethickness{0.6mm}
\put(2,7.5){\line(1,0){5.5}}
\end{picture}  }
\put(100,40){
\unitlength1.mm
\begin{picture}(15,15)
\linethickness{0.15mm}
\put( 6,7.5){\line(1,0){1}}
\put( 4,7.5){\line(1,0){1}}
\put( 2,7.5){\line(1,0){1}}
\put(7.5,2){\line(0,1){5.5}}
\put(7.5,7.5){\line(1,0){5.5}}
\linethickness{0.6mm}
\put(7.5,7.5){\line(0,1){5.5}}
\end{picture}  }
\put(120,40){
\unitlength1.mm
\begin{picture}(15,15)
\linethickness{0.15mm}
\put(7.5, 6){\line(0,1){1}}
\put(7.5, 4){\line(0,1){1}}
\put(7.5, 2){\line(0,1){1}}
\put(2,7.5){\line(1,0){5.5}}
\put(7.5,7.5){\line(0,1){5.5}}
\linethickness{0.6mm}
\put(7.5,7.5){\line(1,0){5.5}}
\end{picture}  }
\put(0,10){
\unitlength1.mm
\begin{picture}(15,15)
\linethickness{0.15mm}
\put(12,7.5){\line(1,0){1}}
\put(10,7.5){\line(1,0){1}}
\put( 8,7.5){\line(1,0){1}}
\put(7.5,7.5){\line(0,1){5.5}}
\put(2,7.5){\line(1,0){5.5}}
\linethickness{0.6mm}
\put(7.5,2){\line(0,1){5.5}}
\end{picture}  }
\put(20,10){
\unitlength1.mm
\begin{picture}(15,15)
\linethickness{0.15mm}
\put(7.5, 6){\line(0,1){1}}
\put(7.5, 4){\line(0,1){1}}
\put(7.5, 2){\line(0,1){1}}
\put(7.5,7.5){\line(1,0){5.5}}
\put(7.5,7.5){\line(0,1){5.5}}
\linethickness{0.6mm}
\put(2,7.5){\line(1,0){5.5}}
\end{picture}  }
\put(40,10){
\unitlength1.mm
\begin{picture}(15,15)
\linethickness{0.15mm}
\put( 6,7.5){\line(1,0){1}}
\put( 4,7.5){\line(1,0){1}}
\put( 2,7.5){\line(1,0){1}}
\put(7.5,7.5){\line(0,1){5.5}}
\put(7.5,7.5){\line(1,0){5.5}}
\linethickness{0.6mm}
\put(7.5,2){\line(0,1){5.5}}
\end{picture}  }
\put(60,10){
\unitlength1.mm
\begin{picture}(15,15)
\linethickness{0.15mm}
\put(12,7.5){\line(1,0){1}}
\put(10,7.5){\line(1,0){1}}
\put( 8,7.5){\line(1,0){1}}
\put(7.5,2){\line(0,1){5.5}}
\put(2,7.5){\line(1,0){5.5}}
\linethickness{0.6mm}
\put(7.5,7.5){\line(0,1){5.5}}
\end{picture}  }
\put(80,10){
\unitlength1.mm
\begin{picture}(15,15)
\linethickness{0.15mm}
\put(7.5,12){\line(0,1){1}}
\put(7.5,10){\line(0,1){1}}
\put(7.5, 8){\line(0,1){1}}
\put(2,7.5){\line(1,0){5.5}}
\put(7.5,2){\line(0,1){5.5}}
\linethickness{0.6mm}
\put(7.5,7.5){\line(1,0){5.5}}
\end{picture}  }
\put(100,10){
\unitlength1.mm
\begin{picture}(15,15)
\linethickness{0.15mm}
\put(2,7.5){\line(1,0){11}}
\put(7.5,2){\line(0,1){11}}
\end{picture} }
\put(0,75){
\parbox[t]{15cm}{
\begin{eqnarray}
\label{allweights}
&&\hspace{-2mm}
\omega_1\hspace{1.6cm}\omega_2\hspace{1.6cm}\omega_3\hspace{1.6cm}
\omega_4\hspace{1.6cm}\omega_5\hspace{1.6cm}\omega_6\hspace{1.6cm}
\omega_7\hspace{3cm}\nonumber\\[2.5cm]
\nopagebreak
&&\hspace{-2mm}
\omega_8\hspace{1.6cm}\omega_9\hspace{1.6cm}\omega_{10}\hspace{1.45cm}
\omega_{11}\hspace{1.45cm}\omega_{12}\hspace{1.45cm}\omega_{13}\hspace{1.45cm}
\omega_{14}\hspace{3cm}\nonumber\\[2.5cm]
\nopagebreak
&&\hspace{-2mm}
\omega_{15}\hspace{1.45cm}\omega_{16}\hspace{1.45cm}
\omega_{17}\hspace{1.45cm}
\omega_{18}\hspace{1.45cm}\omega_{19}\hspace{1.45cm}\omega_{20}\hspace{5cm}
\nonumber
\end{eqnarray}  }}
\end{picture}

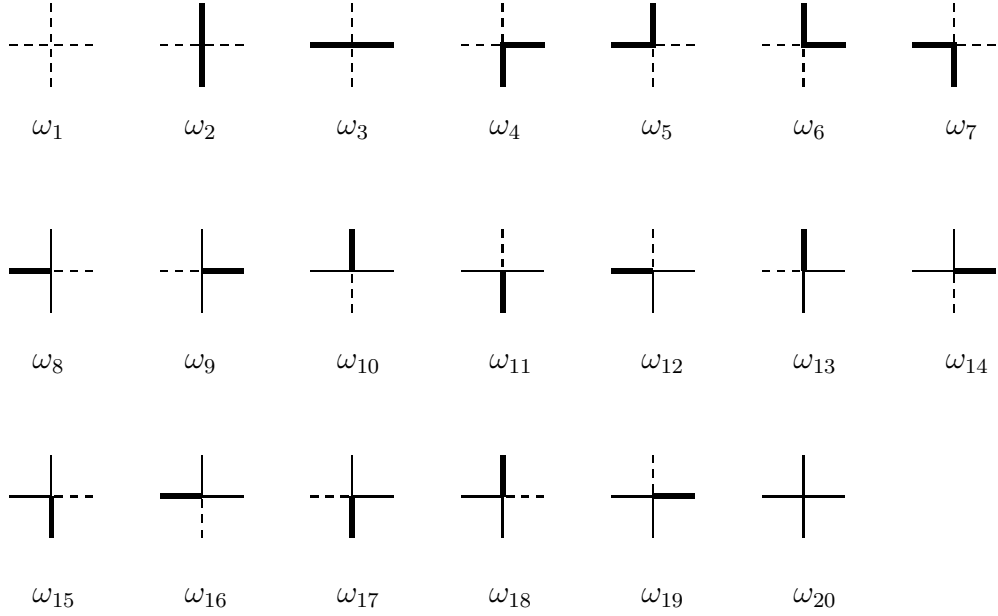
{\bf Figure 1:} Vertices of the (modified) 3-state 20-vertex model on the 
square lattice equivalent to the two-flavour strong coupling Schwinger model
\end{table}

\vspace{5mm}

\begin{eqnarray}
\label{finalweights}
\omega_1&=&z\ =\ {1\over 16\kappa_1^2\kappa_2^2}\ =\ M_1^2 M_2^2\\[3mm]
\omega_2&=&\omega_3\ =\ 1\\[3mm]
\omega_4&=&\omega_5\ =\ \omega_6\ =\ \omega_7\ =\ 
\omega_8\ =\ \omega_9\ =\ \omega_{10}\ =\ \omega_{11}\ =\ 
\omega_{20}\ =\ {1\over 4}\\[3mm]
\omega_{12}&=&\omega_{13}\ =\ \omega_{14}\ =\ \omega_{15}\ =\ 
\omega_{16}\ =\ \omega_{17}\ =\ \omega_{18}\ =\ \omega_{19}\ 
=\ {1\over 2}
\end{eqnarray}

\vspace{5mm}
The partition function reads

\begin{eqnarray}
\label{partition}
Z_\Lambda\ =\ Z_\Lambda[z]&=&\sum_L\ 
z^{\vert\Lambda\vert - \vert L\vert}\ N(L)^2\ 
\left({1\over 2}\right)^{C(L)}\ 
\end{eqnarray}

with

\begin{eqnarray}
\label{partitionadd}
N(L)&=&\sum_{s=0}\ N(s,L)\ 2^s\hspace{1.cm} .
\end{eqnarray}

The sum in eq.\ (\ref{partition}) 
is extended over all possible configurations $L$ 
that can be built from the vertices 1-20. $\vert\Lambda\vert$
is the number of lattice points and $\vert L\vert$ the number
of links with thin or thick lines ($k_l=1,2$). 
$C(L)$ counts the number of 
times a thin line bends (For the purpose of this count thick
lines are understood as being equivalent to two thin lines and the vertices
8-11 therefore represent a twofold bending of a thin line.
For vertex 20, we recall the rule (\ref{rule}).
$P_n(L)$ is the number of times the vertex $n$ appears in 
a given configuration $L$.).

\begin{eqnarray}
C(L)&=&2\ \sum_{n=4}^{11}\ P_n(L)
\ +\ \sum_{n=12}^{19}\ P_n(L)
\ +\ 2\ P_{20}(L)
\end{eqnarray}

The factor $1/2$ then can be understood as a bending 
rigidity of thin lines. The same factor for the bending
rigidity has been found in the one-flavour strong coupling
Schwinger model \cite{salm}.
$N(s,L)$ is the number of configurations with $s$ thin line loops
generated from $L$ by application of the rule (\ref{rule}). 
Following sum rule of course holds.

\begin{eqnarray}
\label{sumrule}
\sum_{s=0}\ N(s,L)&=&2^{P_{20}(L)}
\end{eqnarray}

If $P_{20}(L)=0$, one finds therefore 

\begin{eqnarray}
\label{mult}
N(L)^2&=&4^{S(L)}
\end{eqnarray}

where $S(L)$ denotes the number of thin line loops 
in a given configuration $L$. In this simple case, 
thin line loops enter with a multiplicity of 4 (This case 
bears some similarity to the general loop model
discussed in \cite{rys}.).\\

\parindent1.5em

Eq.\ (\ref{partition}) defines a modified 3-state model
because in the usual definition of a 3-state model the
factor $N(L)^2$ would not be present. Q-state models 
with certain similarities to eq.\ (\ref{partition}) (without the factor $N(L)^2$) have been 
investigated e.g.\
in \cite{strog}-\cite{batch}. For
$z = 0$ we obtain a 19-vertex model. Somewhat similar 
19-vertex models (without the factor $N(L)^2$) have recently
been dicussed in \cite{idzu}-\cite{klue}. Now, the modified
3-state model (\ref{partition}) can alternatively be understood as a 
regular 6-state model. To see this note, that with reference to 
the graphical rules given above each thin line ($k_l=1$) can be
considered as having 4 different states. According to eqs.\ 
(\ref{F1})--(\ref{F8}) each thin line term 
considered within the $\bar\psi$, $\bar\chi$ and $\psi$, $\chi$
subsystems can be related to an arrow assigned to this line. 
This assignment leads to four different states for thin lines
and transforms the 3-state model into a 6-state model. 
Consequently, in this picture the vertices 8-19  
split into four different 6-state model vertices (each corresponds 
to one of the four different products of fields related to 
a given 3-state model vertex, cf.\ eqs.\ (\ref{V18})--(\ref{V16})). 
Each of these vertices has the weight of the corresponding
3-state model vertex.
The vertex 20 yields $6\times 6 = 36$ different 6-state model
vertices (in each subsystem the vertex 20 corresponds to 
6 different products of fields only, cf.\ eq.\ (\ref{V8})). 
The weights of the 6-state model vertices derived from the vertex
20 can easily be determined by expanding the r.h.s.\ of eq.\ (\ref{V8})
in terms of products of fields. Taking into account the above 
considerations one may conclude that the modified 3-state 20-vertex
model (\ref{partition}) is equivalent to a regular 6-state 91-vertex
model. To which extent this is actually useful information
remains to be seen in the future. Another line of thought might be linked 
to the following consideration. As $N(L)$ is related
to some loop count the vertex model (\ref{partition})
bears some resemblance to a loop model. The loop model 
picture becomes even more obvious if one interprets thick 
lines in the present model as two parallel thin lines. If one then 
applies the rule (\ref{rule}) to both the $\bar\psi$, $\bar\chi$ and 
the  $\psi$, $\chi$ subsystems and assigns each subsystem a 
colour, we end up with a 2-colour loop system on the 
square lattice. A loop model which comes closest to this 
picture has been discussed in \cite{war}.
However, the relation to the present model does not go beyond a certain 
remote similarity.\\

Finally, in analogy with the one-flavour case let us discuss 
the critical behaviour of the two-flavour strong coupling 
Schwinger model by means of some approximate method.
We are going to rely on a generalization of the independent
loop approximation (cf.\ sect.\ 3 of \cite{scharn}) which we 
now appropriately call independent cluster approximation. 
As in the case of the independent loop approximation we 
assume that the system under consideration is sufficiently 
dilute in the relevant parameter region in order to 
yield reasonable results. We start with
the observation that any cluster built of the vertices 2-20
falls into two classes. Either we have a self-avoiding thick
line loop or some cluster containing thin line segments. In the 
independent cluster approximation we now write the partition 
function $Z_\Lambda[z]$ as follows.

\parindent0.em

\begin{eqnarray}
\label{cluster}
Z_\Lambda[z]&=&z^{\vert\Lambda\vert}\ 
{\rm e}^{\displaystyle\ \left\{Z_\Lambda[1,z,1/4]\ +\ 
\bar Z_\Lambda[1,z]  \right\}}
\end{eqnarray}

Here, $Z_\Lambda[1,z,\eta]$ is the single loop
partition function defined in sect.\ 3 of \cite{scharn}.
$\bar Z_\Lambda[1,z]$ denotes the single cluster
partition function in which the statistical sum is extended
only over all possible configurations $L$ containing just one cluster
containing thin lines. $z^{\vert\Lambda\vert}\ 
Z_\Lambda[1,z,1/4]$ coincides with $Z_\Lambda[z]$
when the summation in eq.\ (\ref{partition}) is restricted to 
configurations $L$ with just one thick line loop.
$z^{\vert\Lambda\vert}\ 
\bar Z_\Lambda[1,z]$ coincides with $Z_\Lambda[z]$
when the summation in eq.\ (\ref{partition}) is restricted to 
configurations $L$ which constitute just one cluster with
thin line elements. 
As the free energy density $f$ reads in the independent
cluster approximation ($\beta_T = 1/T$)

\begin{eqnarray}
\label{free}
\beta_T f(z)&=& - \lim_{\vert\Lambda\vert\longrightarrow\infty}\ 
{1\over\vert\Lambda\vert}\ \ln Z_\Lambda[z]\nonumber\\[3mm]
&=& \ -\ 
\ln z\ -\ \lim_{\vert\Lambda\vert\longrightarrow\infty}\ 
{ Z_\Lambda[1,z,1/4]\ +\ \bar Z_\Lambda[1,z] 
\over\vert\Lambda\vert}
\end{eqnarray}

a phase transition can be detected by looking at 
each single cluster (loop) partition function separately.
We do not have any information about $\bar Z_\Lambda[1,z]$
but the single loop partition function $Z_\Lambda[1,z,\eta]$
has been studied in \cite{scharn}. One finds a critical line 
described by the equation

\begin{eqnarray}
\label{crit}
z(\eta)&=&1\ +\ \eta\ (\mu-1)\ \ \ \ .
\end{eqnarray}

$\mu = 2.638...$ is the effective coordination number of the 
square lattice. From eq.\ (\ref{crit}) one immediately obtains
in the present case

\begin{eqnarray}
\label{crittwo}
z_{cr}&=&M_1^2 M_2^2\ =\ {1\over 16\kappa_1^2\kappa_2^2}\ =\ 
{(\mu+3)\over 4}\ \ \ .
\end{eqnarray}

 From the experience with the one-flavour \cite{salm} and 
the two-flavour strong coupling Schwinger models it seems 
reasonable to expect that for the investigation of the 
critical behaviour of the general $N_f$-flavour model
the self-avoiding loop model with bending rigidity 
$\eta=2^{-N_f}$ will be relevant.\\

\parindent1.5em

To conclude, we have found the vertex model equivalent to 
the two-flavour strong (infinite) coupling lattice Schwinger model with
Wilson fermions. This demonstrates that the method employed 
by Salmhofer for the one-flavour case \cite{salm} can successfully be
generalized (with some technical effort)
to models with a higher number of Grassmann
variables per lattice site (8 in the present case; incidentally
note that one-flavour strong coupling lattice QED with Wilson fermions
in 4D has the same number of Grassmann variables per lattice 
site). As the method can also be applied to purely 
fermionic models this suggests a number of interesting 
directions for further research which will be explored in the
future.\\

\noindent
{\bf Acknowledgements}\\

The present work has been performed under the EC Human 
Capital and Mobility Program, contract no.\ ERBCHBGCT930470.
I would like to thank Simon Hands for discussions and a critical 
reading of the draft version of the paper, and 
Holger Perlt for valuable advice concerning Mathematica.\\

\end{document}